\documentclass{aa}  

\usepackage{graphicx}
\usepackage{txfonts}
\usepackage[breaklinks=true]{hyperref}
\usepackage{natbib} 
\bibpunct{(}{)}{;}{a}{}{,} 

\usepackage{verbatim}
\usepackage{url}
\usepackage{epstopdf}

\begin{document}

   \title{Grid-based estimates of stellar ages in binary systems}

   \subtitle{SCEPtER: Stellar CharactEristics Pisa
Estimation gRid}

   \author{G. Valle \inst{1,2,3}, M. Dell'Omodarme \inst{3}, P.G. Prada Moroni
     \inst{2,3}, S. Degl'Innocenti \inst{2,3} 
          }

   \authorrunning{Valle, G. et al.}

   \institute{
INAF - Osservatorio Astronomico di Collurania, Via Maggini, I-64100, Teramo, Italy 
\and
 INFN,
 Sezione di Pisa, Largo Pontecorvo 3, I-56127, Pisa, Italy
\and
Dipartimento di Fisica ``Enrico Fermi'',
Universit\`a di Pisa, Largo Pontecorvo 3, I-56127, Pisa, Italy
 }

   \offprints{G. Valle, valle@df.unipi.it}

   \date{Received 22/12/2014; accepted 28/04/2015}

  \abstract
   {}
   {  
We investigate the performance of grid-based techniques in estimating the age of stars in
 detached eclipsing binary systems. We evaluate the precision of the estimates due
to the uncertainty in the observational constraints, $M_{1,2}, T_{eff,1,2}, [Fe/H]_{1,2}$, and the systematic bias caused by the uncertainty in convective core overshooting, element diffusion, mixing-length value, and initial helium content.}
{  
We adopted the SCEPtER grid, which includes stars
with mass in the range 
[0.8; 1.6] $M_{\sun}$ and evolutionary 
stages from the zero-age main sequence to the central hydrogen depletion. Age estimates have been obtained by a generalisation of the maximum
likelihood technique described in our previous work.}
  {
We showed that the typical $1 \sigma$ random error in age estimates -- due only to the uncertainty affecting the observational constraints -- is about $\pm 7\%$, which is nearly independent of the masses of the two stars. However, 
such an error strongly depends on the evolutionary phase and becomes larger and asymmetric for stars near the ZAMS where it ranges from about $+90\%$ to $-25\%$. 
The systematic bias due to the including convective core overshooting -- for mild and strong overshooting scenarios --  is about 50\% and 120\% of the error due to observational uncertainties. 
A variation of  $\pm 1$ in the helium-to-metal enrichment ratio
$\Delta Y/\Delta Z$ accounts for about $\pm 150\%$ of the random error. The neglect of microscopic diffusion accounts for a bias of about 60\% of the error due to observational uncertainties.
We also introduced a statistical test of the expected difference in the recovered age of two coeval stars in a binary system. We find that random fluctuations within the 
current observational uncertainties can lead genuine coeval binary components to appear to be non-coeval with a difference in age as high as 60\%.} 
{}


   \maketitle

\section{Introduction}\label{sec:intro}

The possibility of obtaining direct measurements of stellar mass and radius makes
eclipsing binary systems the objects of fundamental importance \citep[see
  e.g.][]{Andersen1991,Torres2010}.
Nowadays, for stars in double-lined systems, these quantities can be
determined with precisions of a few percentage points or better \citep[see
  e.g.][]{Clausen2008,Pavlovski2014}. Thus, detached
eclipsing binaries offer a unique test bed for evolutionary models of single
stars, giving the opportunity of obtaining accurate estimates of their
  ages. 

However, this age estimation
is usually performed by fitting the
observed data with isochrones or
stellar evolutionary tracks \citep[see
  among many others][]{Southworth2011,Torres2012,Vos2012,Pavlovski2014}.
Recently, different  techniques have been established for single stars, based
upon a maximum 
likelihood estimation over a grid of pre-computed models \citep[e.g.][]{Gai2011,
  Basu2012, Gennaro2012, Mathur2012, PradaMoroni2012, scepter1, eta}. Such methods are very fast and
flexible 
since they can simultaneously take  all the available observables into
account, reaching high
precision in the estimates and  self-consistently evaluating their errors.  

Despite these appealing features, very few efforts can be found in the
literature that address the performance of these techniques for binary systems
and that investigate the possible sources of their internal bias. Two studies that address, among other topics, the problem of binary ages
  are by \citet{Gennaro2012} and \citet{Schneider2014}.
The former is a Bayesian investigation of the age reconstruction of pre-main
sequence stars, which also considers binary systems. The authors adopted nine synthetic
test systems, subjected each of them to 100 Monte Carlo perturbations, and
evaluated the fraction 
of fake non-coeval estimates at $1 \sigma$ level. The results indicated a
fraction of erroneous non-coevality from 0\% to 95\% with a mean of 13\%.
Given the restricted sample of explored systems, \citet{Gennaro2012} recognised the importance of a full analysis of the expected differences in
age estimates in different ranges of mass, metallicity, and evolutionary phases.
The last was a study by
\citet{Schneider2014}, who tested a Bayesian technique for age estimation and,
st other tasks, 
compared the inferred stellar ages for the primary and secondary stars of
eclipsing binaries in the mass range between 4.5 and 28 $M_{\sun}$.
In a sample of 18 binary systems, they found agreement at 1 $\sigma$ level
in the estimated ages of 17 of them. They also evaluated the mean difference in
age of 14 systems (systems with mass ratio higher than 0.97 were excluded) to
be $0.9 \pm 2.3$ Myr at the 95\% confidence level.

The aim of this paper is to theoretically analyse the accuracy of grid-based
age estimates for detached eclipsing binary systems. We also study the bias on
these estimates due to the uncertainties in some physical mechanisms adopted
in the models (convective core overshooting and element microscopic diffusion
efficiencies, mixing-length value), and in the stellar chemical composition
(mainly the initial helium content).  

Finally, we introduce a statistical test of the
expected difference in the reconstructed ages of two coeval binary components
 simply due to the observational uncertainties.  Such a test is of interest
because  the recovered non-coevality of stars in binary systems -- i.e. the inability to fit both components with a single isochrone -- is often used
to claim some deficiency in current generation of stellar models. This sometimes 
leads to introducing and/or calibrating some physical processes in
evolutionary codes, such as convective core overshooting, or to varying the external convection efficiency through the mixing-length parameter
\citep[see e.g.][and references therein]{Andersen1991, Pols1997, 
  Ribas2000a, Torres2006, Morales2009, Clausen2009, Clausen2010, Torres2010, Torres2014}.
 
We restrict our analysis to central hydrogen-burning stars with
mass in the range from 0.8 to 1.6 $M_{\sun}$. A similar
investigation has already been performed \citep{eta} for the age determination of isolated single 
stars for which the main observables adopted in the reconstruction were different from the present ones. 
In that case neither the mass nor the radius were available, while in the present study we do not 
rely on the asteroseismic observables. 
As discussed in the following, the performances and the systematic biases 
of grid-based recovery techniques sensitively depend on the specific set of observables 
actually adopted in the recovery procedure. 

The structure of the paper is the following. In Sect.~\ref{sec:method} we 
discuss the method and the grids used in the estimation process. 
The main results are presented in Sects.~\ref{sec:results} and \ref{sec:propag}.
In Sect.~\ref{sec:expected-diff} we present a statistical test of the expected
differences between the single-star age estimates.  
Some concluding remarks can be found in Sect.~\ref{sec:conclusions}.
The Appendix~\ref{app:sampling} describes the sampling strategy followed 
to build the synthetic dataset.  

\section{Grid-based recovery technique}\label{sec:method}

The basic technique, derived from 
 \citet{Basu2012} and suitable for isolated star estimation, is
 described in \citet{scepter1,eta}, hereafter V14 and V15.
Age estimates are based upon a modified SCEPtER (Stellar CharactEristics Pisa
Estimation gRid) scheme, adapted for binary
stars.  The code and the standard grid developed for this
work are available in the R package
{\it SCEPtERbinary}\footnote{\url{http://CRAN.R-project.org/package=SCEPtERbinary}} on CRAN. 

The  adopted implementation assumes that
 ${\cal S}_1$ and ${\cal S}_2$ are detached binary system stars for which
the 
 following vectors of observed quantities 
are available: $q^{{\cal S}_{1,2}} \equiv \{T_{\rm eff, {\cal S}_{1,2}}, {\rm
  [Fe/H]}_{{\cal S}_{1,2}}, 
M_{{\cal S}_{1,2}}, R_{{\cal S}_{1,2}}\}$. We let $\sigma^{1,2} = \{\sigma(T_{\rm
  eff, {\cal S}_{1,2}}), \sigma({\rm [Fe/H]}_{{\cal S}_{1,2}}), \sigma(M_{{\cal S}_{1,2}}),
\sigma(R_{{\cal S}_{1,2}})\}$ be the nominal uncertainty in the observed
quantities. For each point $j$ on the estimation grid of stellar models, 
we define $q^{j} \equiv \{T_{{\rm eff}, j}, {\rm [Fe/H]}_{j}, M_{j},
R_{j}\}$. 
Let $ {{\cal L}^{1,2}}_j $ be the single-star likelihood functions defined as
\begin{equation}
{{\cal L}^{1,2}}_j = \left( \prod_{i=1}^4 \frac{1}{\sqrt{2 \pi}
  \sigma^{1,2}_i} \right) 
\times \exp \left( -\frac{\chi_{1,2}^2}{2} \right)
\label{eq:lik}
,\end{equation}
where
\begin{equation}
\chi^2_{1,2} = \sum_{i=1}^4 \left( \frac{q_i^{{\cal S}_{1,2}} -
  q_i^j}{\sigma_i} \right)^2. 
\end{equation}

The single-star likelihood functions are  independently evaluated for the two
stars 
for each grid point within $3 \sigma$ of all the variables from ${\cal S}_{1,2}$; let
${{\cal L}^{1,2}}_{\rm max}$ be the two maximum values obtained in this
step. The single-star ages are estimated by averaging the
corresponding quantity of all the models with likelihood greater than $0.95
\times {\cal L}^{1,2}_{\rm max}$.  Informative priors can be inserted as a
multiplicative factor in Eq.~(\ref{eq:lik}), as a weight attached to the grid
points.

In the following we assume, implicitly or explicitly, that the stars in
the binary system are coeval. This assumption justifies the adoption of a single age as representative of the whole system. A first possible way to estimate the age of the binary system 
is simply to take the mean of the ages of the single components,
computed independently. This method has a computational complexity
of $O(n_1 + n_2)$, where $n_{1,2}$ are the sizes of the $3 \sigma$ grid-point
samples.

A second approach explicitly assumes in the likelihood
computation that the 
stars are coeval and computes the joint 
likelihood function only for the couples of models in the two $ 3\sigma$ boxes
with ages  within 10 Myr. The joint likelihood is computed as the
product of the single-star likelihood functions.
Let ${\cal \tilde L}_{\rm max}$ be the
maximum value obtained in this step. The joint-star estimated age is
obtained by averaging the corresponding quantity of all the couples of models
with likelihood greater than $0.95 \times {\cal \tilde L}_{\rm max}$. This
technique has a computational complexity of $O(n_1 \times n_2)$. Readers
interested in the technical solutions implemented are referred 
to the code provided in the above-mentioned {\it SCEPtERbinary} R package.
This estimation method is assumed as
our standard in the following.

The described technique is similar to those previously employed in
the literature in the framework of isochrone
$\chi^2$ fitting for binary age estimations \citep[see, amongst
  many,][]{Pols1997,Lastennet2002}. The main differences are that the quoted
$\chi^2$ approaches adopt a time-consuming numerical functional minimisation  as
the best  
model estimates and use $\chi^2$ profiles to infer the errors on the
estimated parameters, often fixing the value of the masses to their observed
values. However these methods can lead to possibly severe error underestimation
\citep[see e.g.][]{Basu2010, Quirion2010}.
The technique adopted in this and similar grid-based 
approaches \citep[e.g.][]{ Gai2011, Basu2012} computes as best estimate 
a local mean of the
best-matching models, as described above. The error on the estimates
are then computed by 
a Monte Carlo simulation. This approach allows a 
confidence interval to be provided for the age estimate and 
observational parameter correlations to be considered that are not addressed in $\chi^2$
profiles \citep[see e.g.][]{Lastennet2002, Basu2010}.

Although in the current paper we are not interested in obtaining a statistical confidence interval for the age of observed systems, the adopted method can be easily used to this purpose. To address this point,
we reside on a generation of a synthetic
sample of  
$n$ binary systems, starting from the observed values and following, for each star, a multivariate normal
distribution with vector of mean $\{q^{{\cal S}_1}, q^{{\cal S}_2}\}$ and
covariance matrix $\Sigma$. 
For the Monte Carlo simulations, a value of $n =
10\,000$ can be adopted since it provides a fair balance between
computation time and accuracy of the results\footnote{The chosen value
  allows a mean relative accuracy of 0.1\%  to be reached on the $1 \sigma$ confidence
  interval.}. The median of the age of the 
$n$ systems, obtained with one of the two methods described above, is taken as
the best estimate of the true values; the 16th and 84th quantiles of the $n$
values are adopted as a $1 \sigma$ confidence interval.  
 
\subsection{Standard stellar model grid}

The standard estimation grid of stellar models is obtained using FRANEC
stellar evolution code \citep{scilla2008, tognelli2011}, in the same
configuration as adopted to compute the Pisa Stellar
Evolution Data Base\footnote{\url{http://astro.df.unipi.it/stellar-models/}} 
for low-mass stars \citep{database2012, stellar}.

The grid consists of 141\,680 points (110 points for 1\,288 evolutionary
tracks), 
corresponding to evolutionary stages from the ZAMS to the central hydrogen
depletion\footnote{The ZAMS and the central hydrogen depletion 
models are defined as the models whose central hydrogen abundance drops below 99\% 
of the initial value and $10^{-30}$, respectively.}. Models are computed for masses in the range [0.80; 1.60]
$M_{\sun}$ with a step of 0.01 $M_{\sun}$. The initial metallicity [Fe/H]
is assumed in the range [$-0.55$;
  0.55] with a step of 0.05 dex. The solar scaled heavy-element mixture by
\citet{AGSS09} is adopted.  The initial helium abundance is obtained using the
linear relation $Y = Y_p+\frac{\Delta Y}{\Delta Z} Z$
with the primordial abundance $Y_p = 0.2485$ from WMAP
\citep{cyburt04,steigman06,peimbert07a,peimbert07b}, and assuming $\Delta
Y/\Delta Z = 2$ \citep{pagel98,jimenez03,gennaro10}. The models are computed
by assuming the solar-scaled mixing-length parameter $\alpha_{\rm
  ml} = 1.74$. The convective core overshooting extension is
not considered.
Further details on the input adopted in the
computations are available in \citet{cefeidi,incertezze1,incertezze2}.

\section{Binary system age estimates: internal accuracy}\label{sec:results}

To evaluate the internal accuracy of grid-based age estimates of binary systems, 
we analysed the ideal case in which stellar models are in perfect agreement with stars. This is the 
most favourable case where the error in the estimated ages arises only from the 
uncertainties affecting the observational constraints used in the recovery procedure. 

To achieve 
such an ideal situation, the 
 first age-estimation test was performed on a synthetic dataset obtained
by sampling $N = 50\,000$ artificial binary systems from the same standard
estimation 
grid of stellar models used in the recovery procedure. To simulate the effect of observational uncertainties, we added 
a Gaussian noise in all the observed quantities. We assumed as standard
deviations 100 K in $T_{\rm eff}$, 0.1 dex in [Fe/H], 1\% in mass, and 0.5\%
in radius. Since the observationally inferred values of a given physical quantity for the two  binary components are correlated, 
correlation coefficients should be used in the covariance matrix $\Sigma$ whenever a realistic noise has to be simulated. In our standard 
scenario, we assumed a correlation of 0.95 between the primary and secondary effective temperatures, 0.95 between the metallicities, 0.8 between the masses, 
and no correlation between the radii. Motivations of these choices and an analysis of the impact on the results of different assumptions of correlations amongst stellar quantities, radii included, 
are presented in Sect.~\ref{sec:perturbation}.

\begin{figure*}
\centering
\includegraphics[height=17cm,angle=-90]{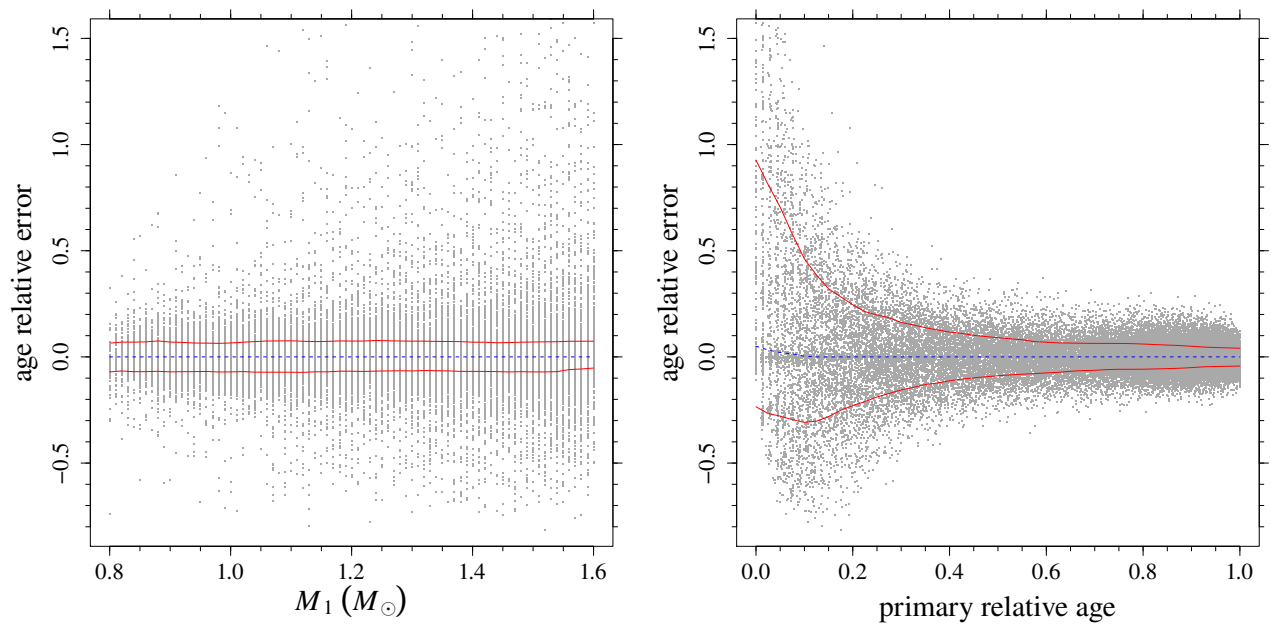}
\caption{{\it Left}: age relative error in
  dependence on the mass of the primary star with its $1
  \sigma$ envelope (red solid line). The blue long dashed line marks the relative errors median.
{\it Right}: same as the {\it left panel}, but in dependence on the relative age of the primary star.}
\label{fig:density-M_pcage-points}
\end{figure*}

Figure~\ref{fig:density-M_pcage-points} shows the Monte Carlo relative error
in estimated age as a 
function of the mass and the relative age of the primary star. In this and in
the 
following analogous figures, a positive 
value of the age relative error corresponds to an overestimated age. 
The relative age is defined as the ratio between
the age of the star and the age of 
the same star at central hydrogen exhaustion (the age is conventionally set to
0 at the ZAMS position).
It shows the position of the median 
relative error and the position of the $1 \sigma$ envelope of age-relative error as a function of
 mass and relative age. The envelope is obtained as in V15,
computing the  
16th and 84th quantiles of the 
relative errors over a moving window in mass and relative age\footnote{The half width
of the windows is chosen to maintain the error on the $1 \sigma$ mass envelope due to Monte
Carlo sampling at a level of about 0.2\%, without introducing too much
smoothing. The corresponding error on the $2 \sigma$ envelope -- relative to
2.5th and 97.5th quantiles -- is about 1\%.}. 

Table~\ref{tab:quantiles} reports the position of the envelope as a function of the mass of the two stars, of their relative age, and of the mass ratio $q$ of the system\footnote{As usual, $q$ is defined as the ratio between the masses of the secondary and the primary stars.}.
Table~\ref{tab:quantiles} and Fig.~\ref{fig:density-M_pcage-points} show that
the typical relative age error 
as a function of the primary star mass 
is about 7\%. The envelope is larger for models near the ZAMS, where the envelope is highly asymmetric 
and -- for the primary star -- it ranges from $-23\%$ to 93\%, with 
a clear edge-effect distortion (see the extensive discussion in V14 and V15). For  relative ages of the primary star larger 
than 0.4 the precision is always better than 12\%.
This accuracy is about four times better than is attainable by adopting
asteroseismic constraints (i.e. the average large frequency spacing $\Delta \nu$ and
the frequency of maximum oscillation power $\nu_{\rm max}$),
without the 
knowledge of mass and radius of the stars (V15 and references therein).
Moreover, in V15 we showed that the systematic in age estimates due to different stellar codes was of the same order 
of magnitude as the random errors. Since the random component is significantly smaller for binary systems, 
we expect the systematic bias due to the adoption of different grids of stellar models to
play a major role in the age determinations.

Several resulting features may depend in principle on the technique employed
to obtain the system age and on the adopted perturbation strategy, so we devoted Sections~\ref{sec:perturbation} and \ref{sec:algorithms} to
discussing the influence of these possible bias sources. 
A detailed discussion of the sampling strategy 
adopted to build the synthetic dataset and an analysis of its effect on the obtained results can be found in 
Appendix~\ref{app:sampling}. 

\begin{table*}[ht]
\centering
\caption{SCEPtER median ($p_{50}$) and $1 \sigma$ envelope boundaries
  ($p_{16}$ and 
  $p_{84}$) for age relative error as a 
  function of the masses, relative ages, and mass ratio of the stars. Values
are expressed as percent.}
\label{tab:quantiles}
\begin{tabular}{rrrrrrrrrrrr}
  \hline\hline
\multicolumn{12}{c}{Mass ($M_{\sun}$)}\\
 & 0.8 & 0.9 & 1.0 & 1.1 & 1.2 & 1.3 & 1.4 & 1.5 & 1.6 &  &  \\ 
  \hline
\multicolumn{12}{c}{primary star}\\
$p_{16}$ & -7.1 & -7.0 & -7.1 & -7.2 & -6.7 & -6.6 & -6.8 & -7.1 & -5.2 &&\\
$p_{50}$ & 0.0 & 0.0 & 0.0 & 0.0 & 0.0 & 0.0 & 0.0 & 0.0 & 0.0 &&\\ 
$p_{84}$ & 6.5 & 7.0 & 6.6 & 7.6 & 7.5 & 7.4 & 7.0 & 7.0 & 7.4 &&\\
 \hline
\multicolumn{12}{c}{Secondary star}\\
$p_{16}$ & -7.2 & -7.0 & -7.1 & -7.1 & -7.1 & -6.7 & -6.3 & -5.3 & -3.4 &&\\
$p_{50}$ & 0.0 & 0.0 & 0.0 & 0.0 & 0.0 & 0.0 & 0.0 & 0.0 & 0.2 &&\\ 
$p_{84}$ & 7.0 & 7.5 & 7.7 & 7.7 & 7.6 & 7.4 & 6.5 & 6.0 & 5.3 && \\ 
\hline\hline
\multicolumn{12}{c}{Relative age}\\
 & 0.0 & 0.1 & 0.2 & 0.3 & 0.4 & 0.5 & 0.6 & 0.7 & 0.8 & 0.9 & 1.0 \\ 
  \hline
\multicolumn{12}{c}{Primary star}\\
$p_{16}$ & -23.5 & -30.8 & -22.9 & -15.5 & -11.3 & -8.9 & -7.5 & -6.3 & -5.8 & -5.0 & -4.3 \\ 
$p_{50}$ & 5.0 & 0.3 & 0.0 & 0.0 & 0.0 & 0.0 & 0.0 & 0.0 & 0.0 & 0.0 & 0.0 \\ 
$p_{84}$ & 92.7 & 46.2 & 24.6 & 16.2 & 11.7 & 9.1 & 6.9 & 6.4 & 6.1 & 5.0 & 4.1 \\ 
 \hline
\multicolumn{12}{c}{Secondary star}\\
$p_{16}$ & -23.3 & -17.0 & -9.8 & -7.7 & -6.8 & -6.3 & -5.8 & -5.2 & -4.8 & -4.3 & -3.9 \\
$p_{50}$ & 0.2 & 0.0 & 0.0 & 0.0 & 0.0 & 0.0 & 0.0 & 0.0 & 0.0 & 0.0 & 0.0 \\ 
$p_{84}$ & 47.3 & 19.3 & 10.3 & 7.9 & 6.5 & 6.3 & 5.8 & 5.3 & 4.9 & 4.3 & 3.9 \\ 
   \hline\hline
\multicolumn{12}{c}{Mass ratio $q$}\\
 & 0.50 & 0.55 & 0.60 & 0.65 & 0.70 & 0.75 & 0.80 & 0.85 & 0.90 & 0.95 & 1.00 \\ 
  \hline
$p_{16}$ &-3.6 & -4.7 & -5.9 & -6.6 & -7.2 & -7.6 & -7.9 & -7.9 & -7.0 & -6.4 & -6.0 \\
$p_{50}$ & 0.0 & 0.0 & 0.0 & 0.0 & 0.0 & 0.0 & 0.0 & 0.0 & 0.0 & 0.0 & 0.0 \\ 
$p_{84}$ & 5.0 & 5.8 & 6.3 & 7.6 & 8.3 & 8.6 & 8.5 & 8.4 & 7.5 & 6.7 & 6.3 \\ 
\hline
\multicolumn{12}{c}{Mass ratio $q$ (alternative sampling)}\\
$p_{16}$ & -5.2 & -5.9 & -6.7 & -7.5 & -8.4 & -9.1 & -9.5 & -9.6 & -8.8 & -8.6 & -8.5 \\
$p_{50}$ & 0.0 & 0.0 & 0.0 & 0.0 & 0.0 & 0.0 & 0.0 & 0.0 & 0.0 & 0.0 & 0.0 \\ 
$p_{84}$ & 4.9 & 6.2 & 7.1 & 8.3 & 9.0 & 9.6 & 10.1 & 10.4 & 9.5 & 9.2 & 8.9 \\ 
\hline
\multicolumn{12}{c}{Mass ratio $q$ (rejection step)}\\
$p_{16}$ &-3.6 & -4.7 & -5.9 & -6.6 & -7.2 & -7.6 & -7.9 & -8.2 & -8.0 & -7.5 & -7.0 \\ 
$p_{50}$ & 0.0 & 0.0 & 0.0 & 0.0 & 0.0 & 0.0 & 0.0 & 0.0 & 0.0 & 0.0 & 0.0 \\ 
$p_{84}$ & 5.0 & 5.8 & 6.3 & 7.6 & 8.3 & 8.6 & 8.5 & 8.6 & 8.6 & 8.2 & 7.9 \\ 
\hline 
\end{tabular}
\end{table*}

\subsection{Effect of correlations}\label{sec:perturbation} 
 
\begin{figure*}
\centering
\includegraphics[height=17cm,angle=-90]{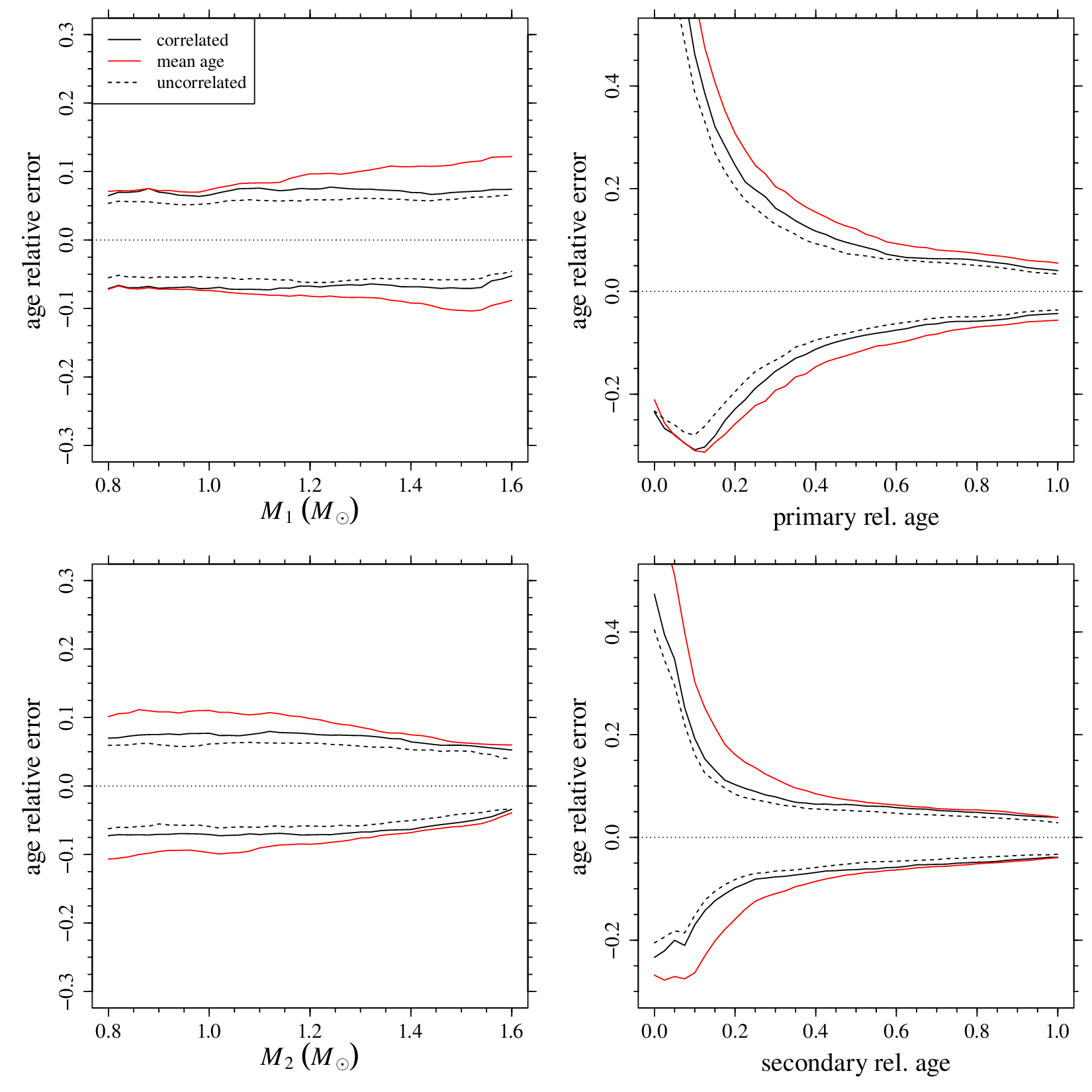}
\caption{{\it Upper row}: $1 \sigma$ envelope of the age relative error in
  dependence on the mass of the primary star ({\it left panel}), and on its
  relative age 
  ({\it right panel}). The solid black line corresponds to accounting for 
  correlations in covariance matrix (see text) 
  and assumes that the stars are coeval; the red solid line assumes the same
  covariance matrix without assumption of coeval stars in the recovery (see Sect.~\ref{sec:method}); the dashed line assumes
  coeval stars and a diagonal covariance matrix.
{\it Lower row}: same as the {\it upper row}, but for the secondary star.}
\label{fig:error}
\end{figure*}

To simulate the observational uncertainties (see Sect.~\ref{sec:method}), we added a Gaussian perturbation 
to the four observables (i.e. $M$, $R$, $T_{\rm eff}$, [Fe/H]) of the stars in the previously generated synthetic dataset of artificial binary systems. In the standard scenario 
they are sampled from a multivariate normal distribution assuming 
correlation of $\rho = 0.95$ between the two effective temperatures, $\rho = 0.95$ between the two metallicities, $\rho = 0.8$ between the two masses, and zero between the two radii.
These correlations arise from the fact that the analysis of data for binary stars 
usually determines some function of the observables at  higher accuracy than the observables themselves.

The strongest correlation is expected between the effective temperatures of the two stars, since the difference in or the ratio of the effective temperatures
 in an eclipsing system is usually determined with relative accuracy that is about three times greater  
 than that of the individual effective temperatures themselves \citep[see e.g.][]{Claret2003,
        Southworth2007, Southworth2013, Torres2014}. 
 A direct computation, by means of error propagation under normality assumption, on the data presented in the cited literature showed that
 the correlation of the two effective 
 temperatures is reproduced well 
 when assuming $\rho = 0.95$.  
 
 Regarding [Fe/H], we adopted a high correlation coefficient ($\rho = 0.95$)
since usually only a mean value for the entire system is presented in the analyses. For mass determination, the mass ratio $q$ is usually determined with relative accuracy 
of about 1.6 higher than the mean one of the masses themselves \citep[see e.g.][]{Popper1986,Lacy2008,Torres2009,Brogaard2011,Vos2012,Sandquist2013}. Assuming normality, this corresponds to a correlation of $\rho = 0.8$ between the two star masses. 
However, even significantly lower values for the correlation ($\rho \approx 0.15$) resulted from literature examination \citep{Helminiak2009, Meibom2009}.  

The situation is much more complex in the case of radii determination. The light curve analysis poses independent constraints on the sum of radii $r_s$ and on their ratio $k$, causing a dependence between the two evaluated stellar radii; however, the magnitude and the sign of the correlation between the estimated radii depend on the actual values of $r_s$ and $k$ and of their errors. Direct computations from the values quoted in the literature showed correlations varying from $\rho = 0.8$ to $\rho = -0.5$ \citep{Popper1986,Grundahl2008,Torres2009,Brogaard2011,Vos2012}.
Luckily, such high variability is not a serious problem because, as shown below, the actual magnitude of the correlation between radii has very little impact on the results. Thus, for 
the reference scenario we assumed uncorrelated radii. 

To quantify the impact on the age estimate uncertainty of accounting for the aforementioned correlations in 
the procedure followed to build the synthetic datasets, we applied the SCEPtER pipeline on datasets produced by relying 
on different assumptions about correlations. As a first test, we provided a dataset of artificial binary stars with uncorrelated observables; 
i.e., the Gaussian noise was added assuming a diagonal covariance matrix. Although such a choice is quite extreme 
and not realistic, the comparison between the outcomes of the age estimate procedure 
on this dataset with those on the standard one will quantify the maximum effect of the correlations.
Figure~\ref{fig:error} shows the impact of the correlation between observables on the
relative errors in the estimated age of the system. 
 The neglect of correlation on the corresponding observables of the two stars
causes a mean relative shrinkage of the envelope of about 20\%. Direct simulations (not shown) proved that the shrinkage
 is due to dropping the correlation between the two masses, while the correlations between the effective temperatures and between the metallicities account for modest variations.

As a second test, we built two synthetic datasets by assuming the same correlations of the standard case, with the exception of the correlation 
coefficient between the radii of the binary components, namely $\rho = \pm 0.7$. 
The comparison of the performances of the recovery procedure on these datasets with those on the standard one will assess 
the impact of the correlation amongst the radii. Table~\ref{tab:radcorr} reports, as a function of the mass of the primary star, the results obtained accounting for correlations between the radii. 
The mean variations on the envelope boundaries are about 0.2\%, so it is safe to neglect the correlation among radii.

\begin{table*}[ht]
\centering
\caption{Median ($p_{50}$) and $1 \sigma$ envelope boundaries
  ($p_{16}$ and 
  $p_{84}$) for age relative error as a function of the mass of the primary
  star, assuming correlations $\rho = \pm 0.7$ between the radii.} 
\label{tab:radcorr}
\begin{tabular}{rrrrrrrrrr}
  \hline\hline
\multicolumn{10}{c}{Primary star mass ($M_{\sun}$)}\\
 & 0.8 & 0.9 & 1.0 & 1.1 & 1.2 & 1.3 & 1.4 & 1.5 & 1.6 \\ 
  \hline
\multicolumn{10}{c}{$\rho = 0.7$}\\
  \hline
$p_{16}$  & -7.1 & -6.9 & -7.3 & -7.4 & -6.9 & -7.2 & -7.0 & -6.8 & -5.2 \\ 
$p_{50}$  & 0.0 & 0.0 & 0.0 & 0.0 & 0.0 & 0.0 & 0.0 & 0.0 & 0.0 \\ 
$p_{84}$  & 6.9 & 6.6 & 6.6 & 7.3 & 7.0 & 7.1 & 6.9 & 7.3 & 7.4 \\ 
\hline
\multicolumn{10}{c}{$\rho = -0.7$}\\
 \hline
$p_{16}$  & -6.4 & -6.6 & -7.0 & -7.2 & -6.8 & -6.8 & -6.8 & -6.9 & -5.3 \\ 
$p_{50}$  & 0.0 & 0.0 & 0.0 & 0.0 & 0.0 & 0.0 & 0.0 & 0.0 & 0.0 \\ 
$p_{84}$  & 7.1 & 7.4 & 6.7 & 7.4 & 7.5 & 7.0 & 7.0 & 7.1 & 7.0 \\ 
   \hline
\end{tabular}
\end{table*}

\subsection{Recovery algorithms}\label{sec:algorithms} 

Figure~\ref{fig:error} also shows  the impact of the adopted age-recovery
technique. The simple average of the individually independent age estimates of the two stars
appears to be as good as the joint estimate in the case of a primary star of low mass ($M \leq 1.1$ $M_{\sun}$), 
while it shows a notable larger variance in binary systems composed of a high-mass primary 
and a low-mass secondary. For these combinations of objects, the
secondary stars is sampled in the first stages of the evolution, since the
maximum time is set by the faster evolution of the more massive star.
The age estimate of a star in the early evolutionary stages is subject
to larger relative uncertainty, and this uncertainty is propagated in the
final mean. Such an effect is not present in the joint evaluation, since in this
case the age is mainly determined by the more evolved star. In fact the
age spread of
the models in the $3 \sigma$ box around the primary star is generally lower
than the spread the secondary star, since the primary is sampled at
a higher relative age than the secondary. Therefore the joint likelihood
estimate is limited by the range of age spanned by the primary star.

\begin{figure*}
\centering
\includegraphics[height=17cm,angle=-90]{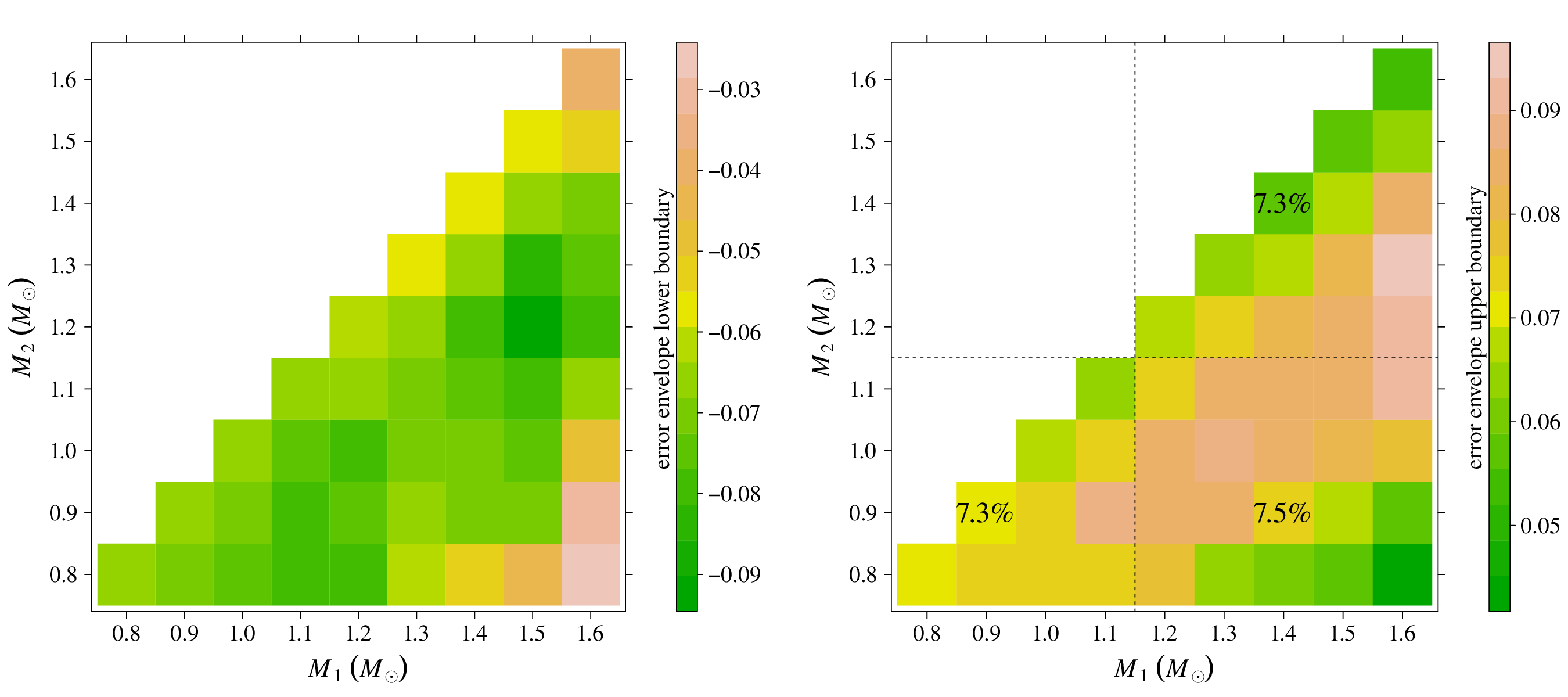}\\
\includegraphics[height=17cm,angle=-90]{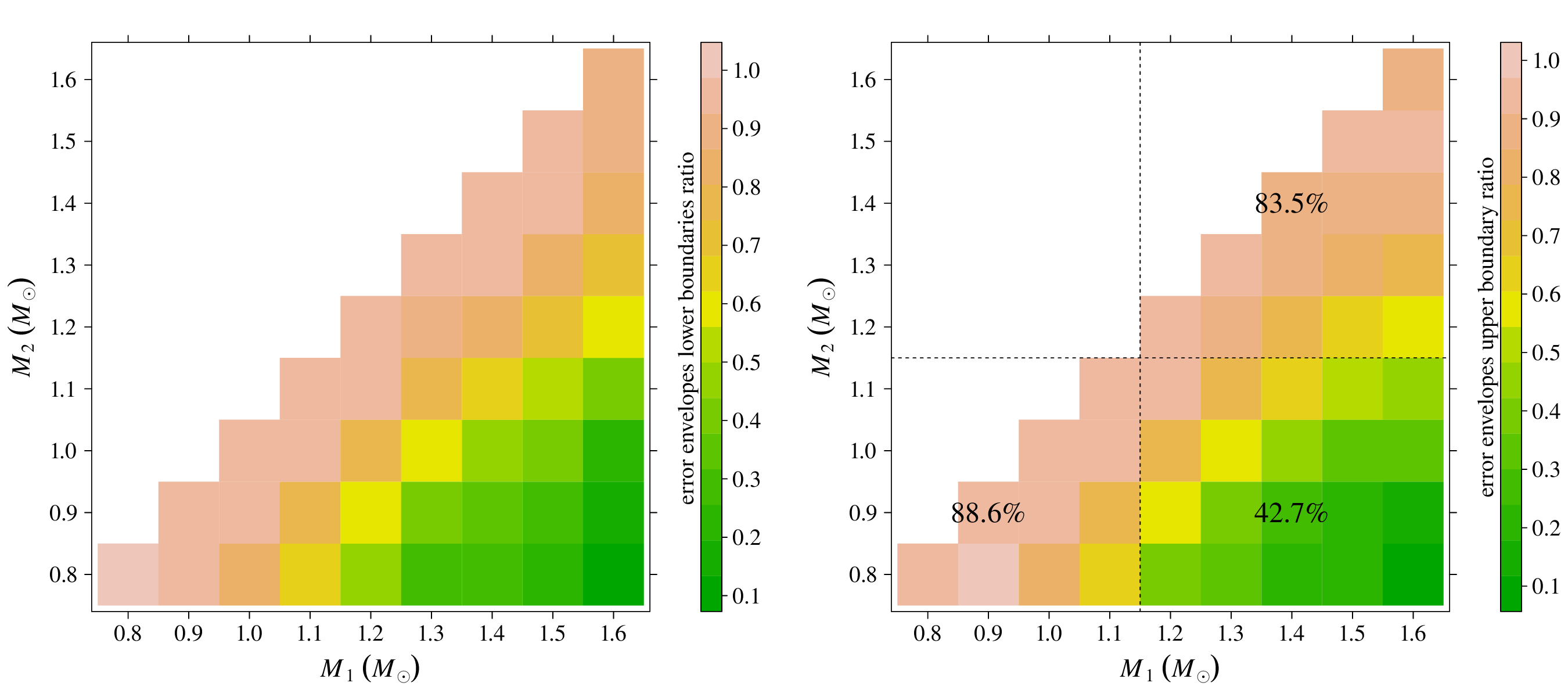}
\caption{{\it Upper row}: {\it left} lower boundary of the $1 \sigma$ 2D
  relative error envelope for 
  joint likelihood estimates as a function of the mass of the binary system
  stars. The percentages on the plot refer to mean values in different ranges
  of mass (see text). {\it Right}: same as the {\it left panel}, but for the
  upper 
  boundary of the 2D relative error envelope.  {\it Lower row}: {\it left} ratio of
    lower 
  boundary of the $1 \sigma$ 2D relative error envelope due to different recovery techniques. The ratio is computed by dividing the position of the joint likelihood estimates boundary by the  corresponding
  value obtained averaging the independently estimated ages.
    {\it Right}: same as the {\it left panel}, but for the upper
  boundary of 
  the 2D relative error envelope.  }
\label{fig:envelop-M1_M2-diff12}
\end{figure*}

The increase in the variance in age estimates when coevality is not explicitly assumed is shown further in Fig.~\ref{fig:envelop-M1_M2-diff12}. The top 
row of the figure shows, as
a function of the masses $M_1$ and $M_2$ of the binary system components, the
 position of the 2D lower (left panel) and upper (right panel) $1 \sigma$
 envelopes\footnote{The bidimensional envelopes are computed with a similar technique
        to the 1D ones, with mass steps of 0.1 $M_{\sun}$ and moving-window half width of
        0.08 $M_{\sun}$. The mass binning choice is such that the envelope is 
        affected by a typical random uncertainty of about 0.5\%.}. 
To summarise the results, the figure reports the mean value (expressed as percent) of the
envelope boundaries in three mass ranges: 
 $M_1$ < 1.1 $M_{\sun}$ and $M_2$ < 1.1 $M_{\sun}$;  $M_1$ > 1.1 $M_{\sun}$ and
 $M_2$ > 1.1 $M_{\sun}$; $M_1$ > 1.1 $M_{\sun}$ and $M_2$ < 1.1 $M_{\sun}$.
The lower row of the figure displays the ratio of the boundaries
for joint likelihood and independent age estimates.   
It is apparent that the two age-estimation techniques give nearly equivalent results as
long as the system stars are close to each other in mass ($ q \approx 1 $), while for unbalanced
systems, the ratio of the boundary is as low as 0.2 for $M_1$ = 1.6 $M_{\sun}$, $M_2$ = 0.8 $M_{\sun}$ , which is, an increase in the envelope width of a factor of five.

\subsection{Contribution of the secondary star to age estimates}\label{sec:secondary}

\begin{figure*}
\centering
\includegraphics[height=17cm,angle=-90]{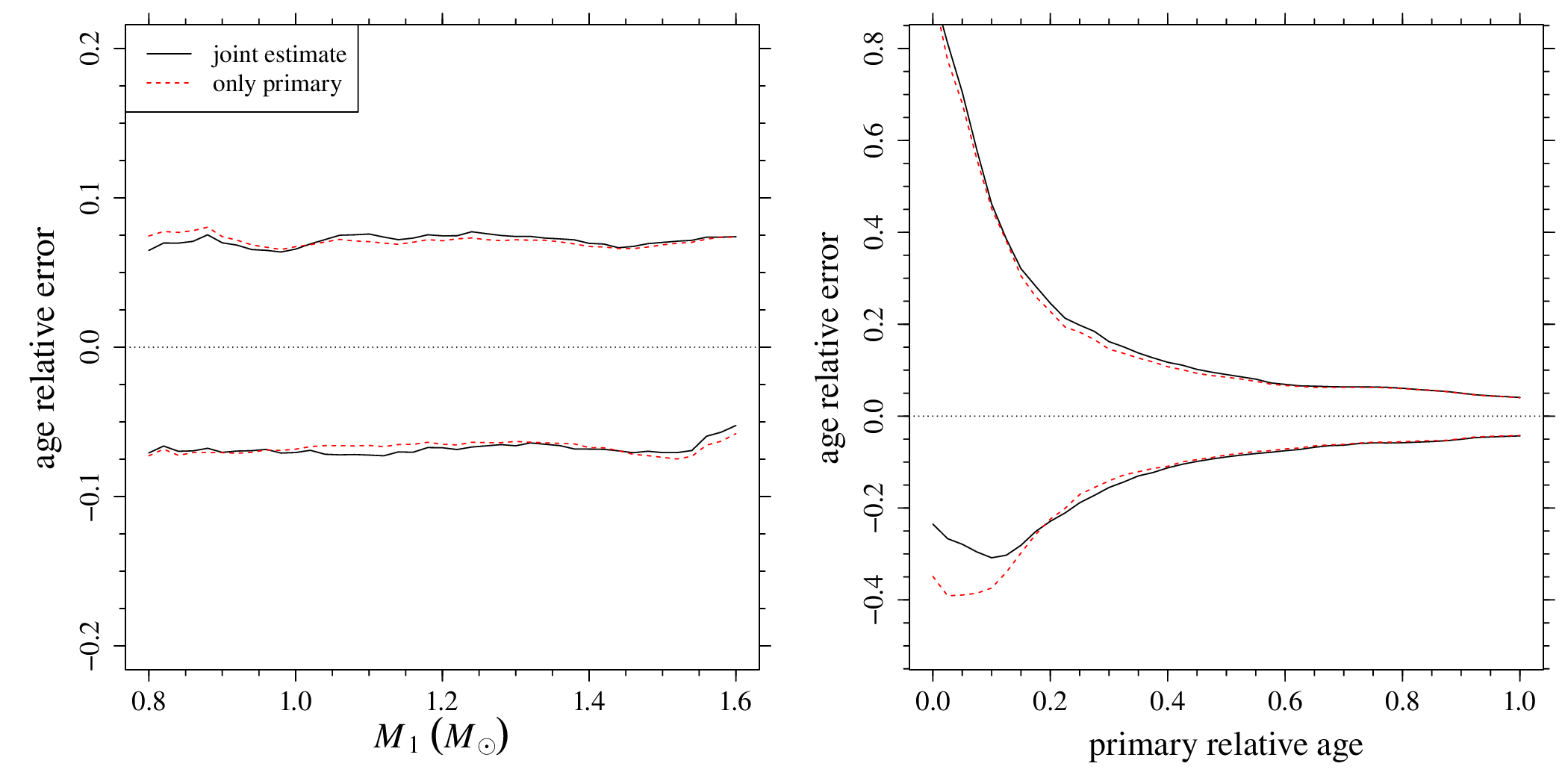}
\caption{{\it Left}: $1 \sigma$ envelope of the age-relative error in
  dependence on the mass of the primary star (solid line), compared with the
  same quantity computed only with the data of the primary star (dashed line).
{\it Right}: same as the {\it left panel}, but as a function of relative age of the primary star.}
\label{fig:error-primary}
\end{figure*}

In Sect.~\ref{sec:algorithms} we discussed the dominant impact of the primary star on
joint likelihood age determination. It is therefore interesting to
explore the contribution of the secondary star to the whole system age
determination. We performed this test by comparing the envelopes for the joint
likelihood age estimate and for the primary single-star estimates. The results
are presented in Fig.~\ref{fig:error-primary}. Computed as a function of the primary mass, the joint
estimate's relative 
error envelope shows negligible differences for both the upper and the lower
boundaries with respect to the envelope computed only for primary stars. The contribution of the secondary star is only visible for
young systems, where the age of the primary star is less constrained. In fact, the right-hand 
panel of the figure shows that the relative age envelope that takes the secondary contribution into
account is less prone to age underestimation at a relative age lower than 0.15. 
In summary, the secondary star plays a minor role with respect to the primary one because its age is generally 
estimated with a larger uncertainty being closer to the ZAMS than the more massive companion of the same age.

\section{Stellar model uncertainty propagation}\label{sec:propag}

The accuracy of age estimates of real binary systems obviously depends on the
reliability of the adopted grid of stellar models. In this section, following V14 and V15, we provide an analysis of the impact of the various
uncertainty sources that affect the stellar computations.
We focus here on some issues considered in V15: the convective core overshooting efficiency, the initial helium
content, and the neglect of the microscopic diffusion.
We constructed several artificial
grids characterised by the variation in the single input under
examination up to its 
extreme allowed values, while keeping all the others fixed to their reference
values. We then built the corresponding synthetic datasets by sampling artificial binary systems from these non-standard grids of stellar models and 
by adding a Gaussian noise in all the observed quantities to simulate observational uncertainties following the same 
prescriptions for covariance matrix described in the standard case. Finally, the ages are estimated by means of the SCEPtER pipeline by relying on the standard 
grid of stellar models.

Although the use of a solar-calibrated mixing-length
value for stars that differ from the Sun for mass, composition, or 
evolutionary phase could be inappropriate in the present 
paper; unlike V15, we do not consider this
source of uncertainty. 
 Such a choice is motivated by the fact that the dependence of the mixing-length
 value on the mass, metallicity, level of activity, and the evolutionary phase of
 the star is still under study and that no 
 definitive conclusion has been reached \citep[see, among many,][]{Clausen2009,
 Deheuvels2011,  Bonaca2012, Mathur2012, Tanner2014}. 
In V15  we studied the impact on age estimates due to a change in the mixing-length value by adopting synthetic grids with non-solar mixing-length values, which did not vary inside a given grid
regardless of the mass and metallicity of the star. Since in the present work, we consider systems of stars with different masses, hence in different evolutionary stages, we feel that the approach of V15 might
 not be appropriate. Besides the effect of a  change in the mixing length of the sampling grid, a differential effect of the possible evolution of the mixing-length value with the mass of the stars could be important.  
Nevertheless, we performed a numerical
experiment, by sampling from grids with $\alpha_{\rm ml} \pm 0.24$ with respect
to our solar-calibrated value ($\alpha_{\rm ml}$ = 1.74), which showed a bias ranging
from about $\mp 4.0\%$ to $\mp 6.0\%$, the lowest values for  mass ratio $q$
near 1.0. These simulations suggest that the bias due to the different
choices of the mixing-length values is certainly important but not dominant.
For a more rigorous investigation, better knowledge of the dependence 
of the external convection efficiency on
 stellar parameters, such as mass, chemical composition, evolutionary phase,
 and stellar activity, is mandatory.

The observables adopted here in the age reconstruction are different from those
in V15, which made use of classical (i.e. the effective temperature of stars and their metallicity [Fe/H]) and asteroseismic (i.e. the average large frequency spacing $\Delta \nu$ and
the frequency of maximum oscillation power $\nu_{\rm max}$) observable constraints.
We therefore expect that the impact of the uncertainty sources is 
different from what is found in our previous work. In particular, the availability of precise stellar masses
 is paramount since it severely restricts
the number of stellar models that can actually contribute to the recovery procedure.  

\begin{table*}[ht]
\centering
\caption{SCEPtER bias on the recovered age due to the change in the stellar code input. Median age
  relative errors ($p^{\rm p}_{50}$ and $p^{\rm s}_{50}$) are reported as a 
  function of the mass of the primary  and secondary  star. Values
are expressed as percent.}
\label{tab:biasM}
\begin{tabular}{rrrrrrrrrr}
  \hline\hline
 & \multicolumn{9}{c}{Mass ($M_{\sun}$)}\\
 & 0.8 & 0.9 & 1.0 & 1.1 & 1.2 & 1.3 & 1.4 & 1.5 & 1.6\\
  \hline
\multicolumn{10}{c}{Overshooting $\beta = 0.2$}\\
$p^{\rm p}_{50}$ &  &  & 0.0 & -0.7 & -3.6 & -6.2 & -5.7 & -6.1 & -4.9 \\
$p^{\rm s}_{50}$ &-0.2 & -0.2 & -1.0 & -3.0 & -4.7 & -6.6 & -6.8 & -6.6 & -6.1 \\
  \hline
\multicolumn{10}{c}{Overshooting $\beta = 0.4$}\\
$p^{\rm p}_{50}$ &   &  & 0.0 & -3.3 & -10.3 & -10.7 & -10.2 & -10.7 & -10.5 \\ 
$p^{\rm s}_{50}$ & -0.8 & -1.6 & -4.0 & -8.2 & -11.9 & -10.9 & -11.2 & -11.2 & -9.4 \\ 
  \hline
\multicolumn{10}{c}{$\Delta Y/\Delta Z = 1$}\\
$p^{\rm p}_{50}$ & -6.4 & -6.8 & -9.7 & -10.1 & -10.7 & -10.0 & -9.8 & -9.2 & -7.8 \\ 
$p^{\rm s}_{50}$ & -7.9 & -8.3 & -10.0 & -10.2 & -9.9 & -9.7 & -9.2 & -8.1 & -6.8 \\ 
  \hline
\multicolumn{10}{c}{$\Delta Y/\Delta Z = 3$}\\
$p^{\rm p}_{50}$ & 6.9 & 10.1 & 10.5 & 11.4 & 11.4 & 10.5 & 10.1 & 9.7 & 10.2 \\
$p^{\rm s}_{50}$ &9.4 & 10.5 & 11.6 & 10.6 & 10.5 & 10.4 & 9.7 & 8.7 & 8.6 \\
  \hline
\multicolumn{10}{c}{No microscopic diffusion}\\
$p^{\rm p}_{50}$ & 8.0 & 6.2 & 5.1 & 4.0 & 3.8 & 3.3 & 2.8 & 2.6 & 2.9 \\
$p^{\rm s}_{50}$ & 6.4 & 4.9 & 3.9 & 3.4 & 3.3 & 3.1 & 2.6 & 2.0 & 1.7 \\
   \hline
\end{tabular}
\end{table*}

\begin{table*}[ht]
\centering
\caption{SCEPtER bias on the recovered age due to the change in the stellar code input. Median age
  relative errors are reported as a 
  function of the mass ratio of the binary system. Values
are expressed as percent.}
\label{tab:biasq}
\begin{tabular}{lrrrrrrrrrrr}
  \hline\hline
 & \multicolumn{11}{c}{Mass ratio $q$}\\
Input &  0.50 & 0.55 & 0.60 & 0.65 & 0.70 & 0.75 & 0.80 & 0.85 & 0.90 & 0.95 & 1.00 \\ 
  \hline
$\beta = 0.2$ & -3.7 & -3.6 & -3.7 & -3.8 & -3.9 & -3.8 & -3.5 & -3.2 & -2.6 & -3.4 & -3.4 \\ 
$\beta = 0.4$ & -8.2 & -9.2 & -9.7 & -9.8 & -9.5 & -9.0 & -8.1 & -7.3 & -6.1 & -5.0 & -4.6 \\ 
$\Delta Y/\Delta Z = 1$ & -7.6 & -8.0 & -9.0 & -9.3 & -9.7 & -9.8 & -9.9 & -9.8 & -8.9 & -9.1 & -8.9 \\ 
$\Delta Y/\Delta Z = 3$ & 10.0 & 10.5 & 11.2 & 11.3 & 11.1 & 11.3 & 11.3 & 11.2 & 10.2 & 9.9 & 9.7 \\ 
No diffusion & 4.0 & 3.8 & 3.4 & 3.3 & 3.4 & 3.7 & 3.9 & 4.1 & 3.8 & 3.4 & 3.2 \\ 
   \hline
\end{tabular}
\end{table*}

\subsection{Convective core overshooting}
\label{sec:ov}

As in V15 we parametrised the extension 
of  the extra-mixing region beyond the canonical border, as defined by the 
Schwarzschild criterion, in terms of the pressure scale height $H_{\rm 
  p}$: $l_{\rm ov} = \beta H_{\rm p}$, where $ \beta $ is a free parameter. 
  To quantify the impact of taking the convective core overshooting into account, we
computed -- only for models more massive than 1.1 $M_{\sun}$ --  two additional
grids with values of $\beta$ = 0.2 and 0.4,  the latter representing a
  possible 
maximum value \citep[see e.g. the discussion
  in][]{cefeidi}.
 We then extended these two grids of stellar models at masses lower than 1.1 $M_{\sun}$ with the standard one, which neglect core overshooting.

\begin{figure*}
\centering
\includegraphics[height=17cm,angle=-90]{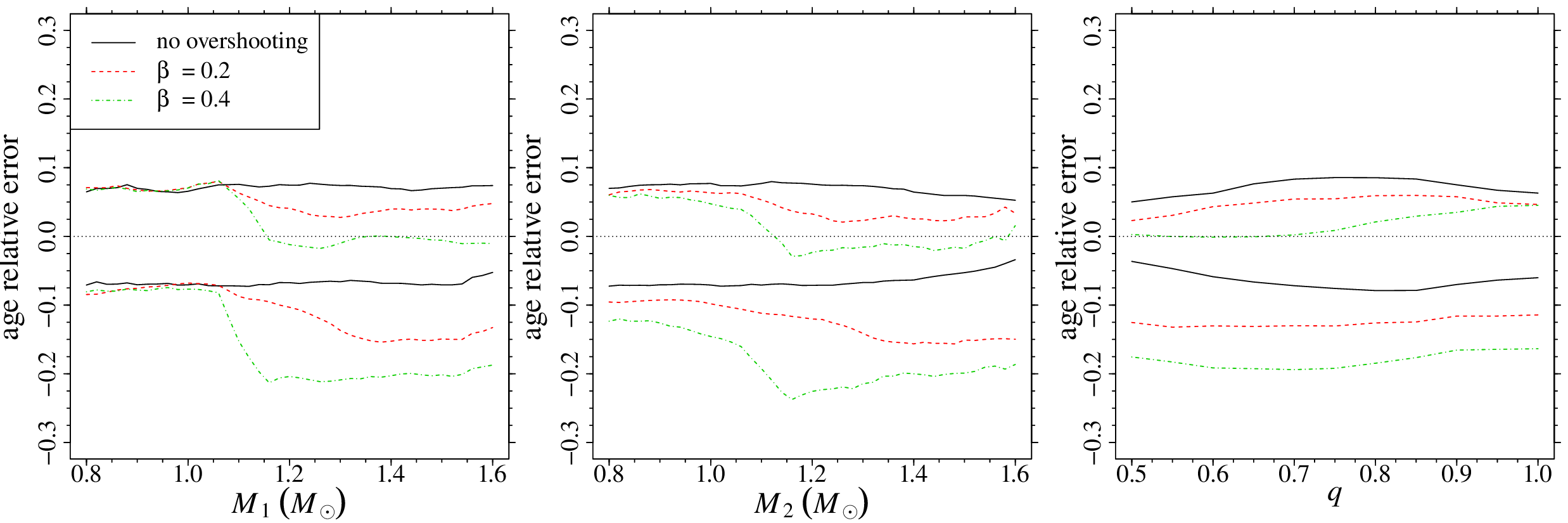}
\caption{{\it Left}: standard $1 \sigma$ envelope of the binary system age
  relative error in 
  dependence on the mass of the primary star (solid black), compared with the
  same quantity obtained by sampling from grids with different core
-overshooting efficiency $\beta = 0.2$ (dashed red) and $\beta = 0.4$ (dot
  dashed green).
{\it Middle}: same as the {\it left panel}, but in dependence on the
mass of the secondary star. {\it Right}: same as the {\it left panel}, but in dependence on the
  system  mass ratio $q$.}
\label{fig:error-OV}
\end{figure*}

\begin{figure*}
\centering
\includegraphics[height=8.5cm,angle=-90]{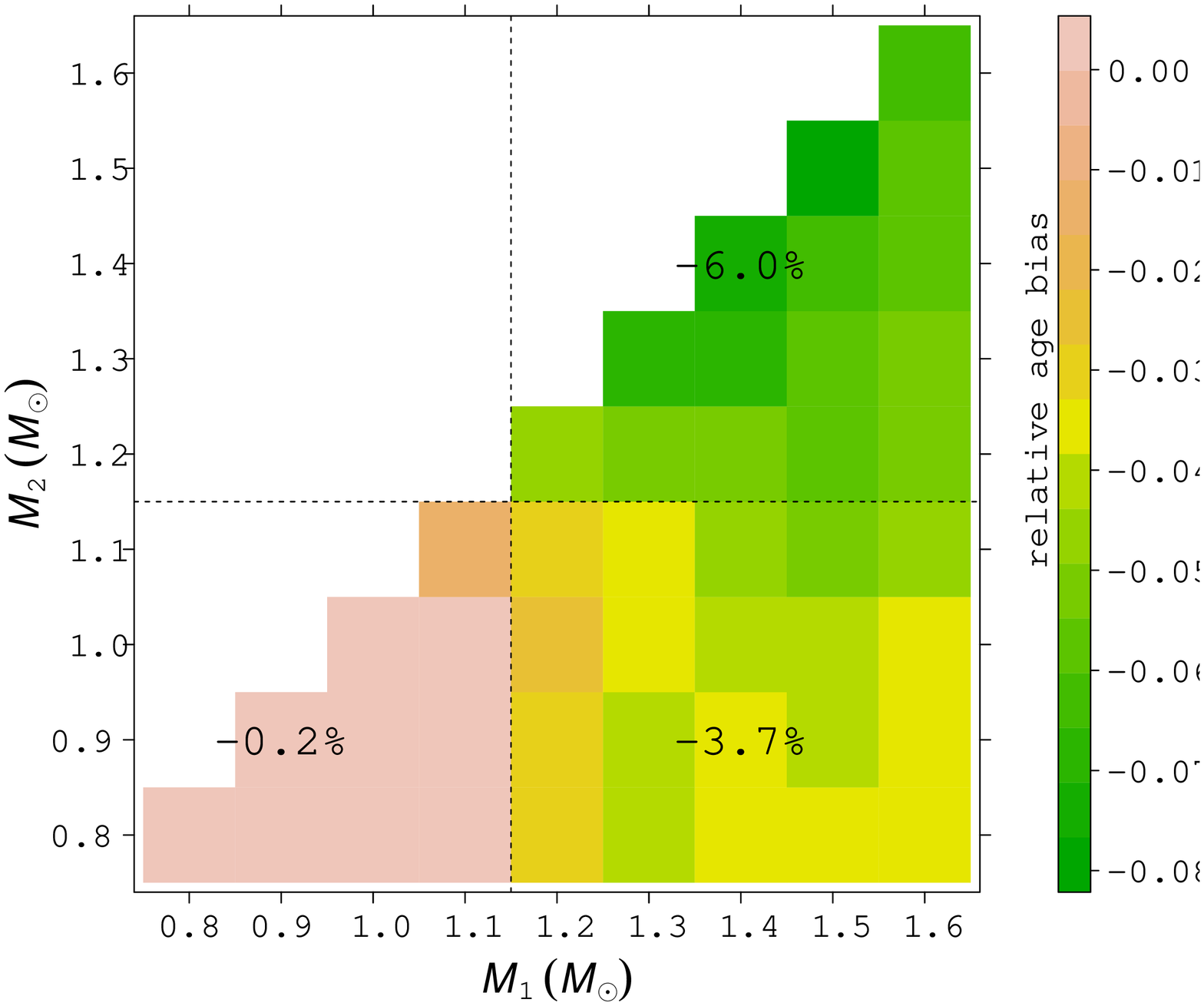}
\includegraphics[height=8.5cm,angle=-90]{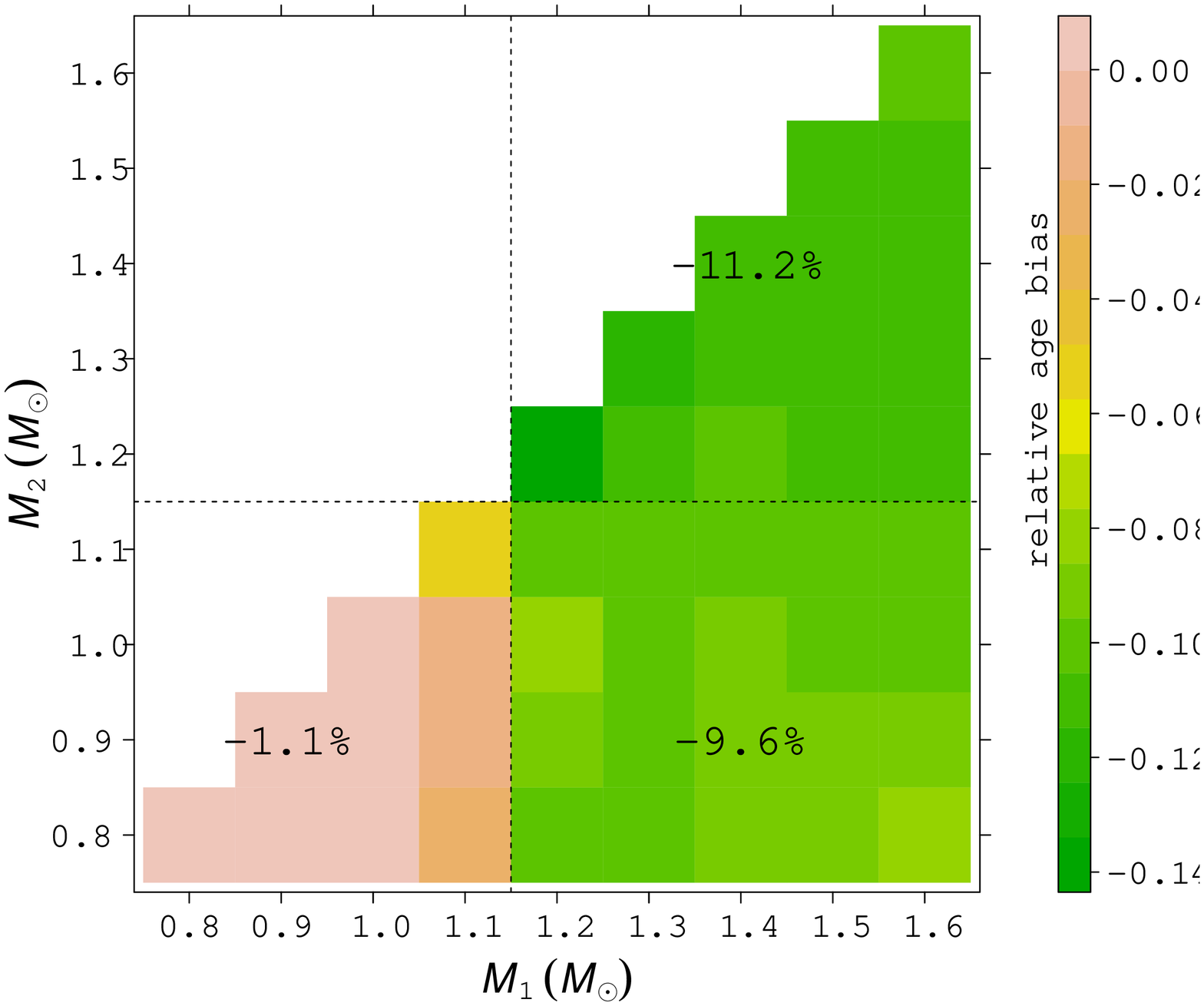}
\caption{{\it Left}: relative age bias due to the mild core overshooting
as a function of the masses of the binary system. Artificial stars      
are sampled from the grid with mild core overshooting ($\beta =
  0.2$) and their age  is estimated on the standard grid. {\it Right}: same as the
  {\it left
  panel}, but for the strong overshooting
scenario 
  ($\beta =  0.4$).}
\label{fig:envelop-M1_M2-diff-OV}
\end{figure*}

From these non-standard grids we built two synthetic data sets (for $\beta = 0.2$ and $\beta = 0.4$) by sampling $N = 50\,000$ 
artificial binary systems each and adding a Gaussian noise. The age estimate is then performed by adopting 
our standard grid of models (without overshooting) in the recovery procedure.  
The results are shown in Tables~\ref{tab:biasM} and \ref{tab:biasq} and in
Figs.~\ref{fig:error-OV} and \ref{fig:envelop-M1_M2-diff-OV}.
To evaluate the magnitude of the obtained biases, they can be compared with the  
standard $1 \sigma$ envelope due to observational uncertainties in Table~\ref{tab:quantiles}. 
It turns out that the median bias due to the mild (i.e. $\beta = 0.2$) core overshooting
scenario -- computed as function of the mass ratio $q$ -- is about 50\%
of the half width of the standard envelope uncertainties. The corresponding bias for the
strong ($\beta = 0.4$) core overshooting scenario is about 120\% of the standard envelope
half width. In both cases the relative importance of the bias is generally greater
for low values of $q$; for the strong overshooting scenario, 
in particular,  the bias reaches
values of about 160\% of the standard envelope half width at $q$ = 0.5, while it is only 60\%
 at $q$ = 1.0. 
This trend is expected since part of the systems with $q$ near 1.0 are
composed of stars with both masses lower than 1.1 $M_{\sun}$, for which 
core overshooting is not taken into account so that the discussed effect is partially masked. 

Figure~\ref{fig:envelop-M1_M2-diff-OV} shows
that the dominant effect for the bias is due to the mass of the primary star. 
As an example, the mean bias for mild overshooting scenario (left panel in the figure)
 is $-0.2\%$ for a primary star that is less massive than 1.1 $M_{\sun}$, while it is $-3.7\%$ for a massive primary coupled to a  light secondary star.
 The shift due to having a massive secondary star is $-6.0\%$, so it only accounts for an additional $-2.3\%$ change. 
 The contribution of the secondary star is even less important in the strong overshooting scenario (right panel in the figure), when it accounts for about one-fifth 
 of the primary star one.

The results presented in this section have particular importance since binary systems are often adopted to calibrate the efficiency of the convective core overshooting by isochrone fine tuning \citep[see e.g.][]{Claret2007,Lacy2008,Clausen2010}. 
Although our results do not directly allow estimation of the errors in the recovered best convective core-overshooting parameter that is consistent with the observations, we found that the bias induced by a mild convective core overshooting scenario on age estimate is  about one-half of the $1 \sigma$ expected error in age estimate. This implies a difficulty unambiguously identifying the overshooting effect, which can be masked  by random fluctuations. 

The present results suggest that a mild overshooting scenario is 
hardly distinguishable from a no overshooting one. This result agrees with the finding by \citet{Claret2007} in the lower mass range considered in that paper.
Although a statistical calibration -- obtained from several systems -- can be meaningful, the large errors that propagate into the final calibrated overshooting parameter
shed some doubt on the calibration of single systems.
Owing to the widespread use of binary systems for calibration purposes, we feel that further research is needed to statistically quantify the errors in the calibrated parameters.

\subsection{Initial helium abundance}
\label{sec:he}

We quantified the impact on the age estimate of the current uncertainty in 
 the initial helium abundance following V15. We computed two additional grids of stellar models with
the same metallicity values $Z$ as in the standard grid, but by
changing the helium-to-metal enrichment ratio $\Delta Y/\Delta Z$ to values of 1
and 3. Then, we built two synthetic datasets, each of $N = 50\,000$ artificial
binary systems, 
by sampling the objects from these two non-standard grids and adding Gaussian noise. The age of the
objects is reconstructed using the SCEPtER pipeline relying on the standard grid with $\Delta Y/\Delta Z = 2$.

\begin{figure*}
\centering
\includegraphics[height=17cm,angle=-90]{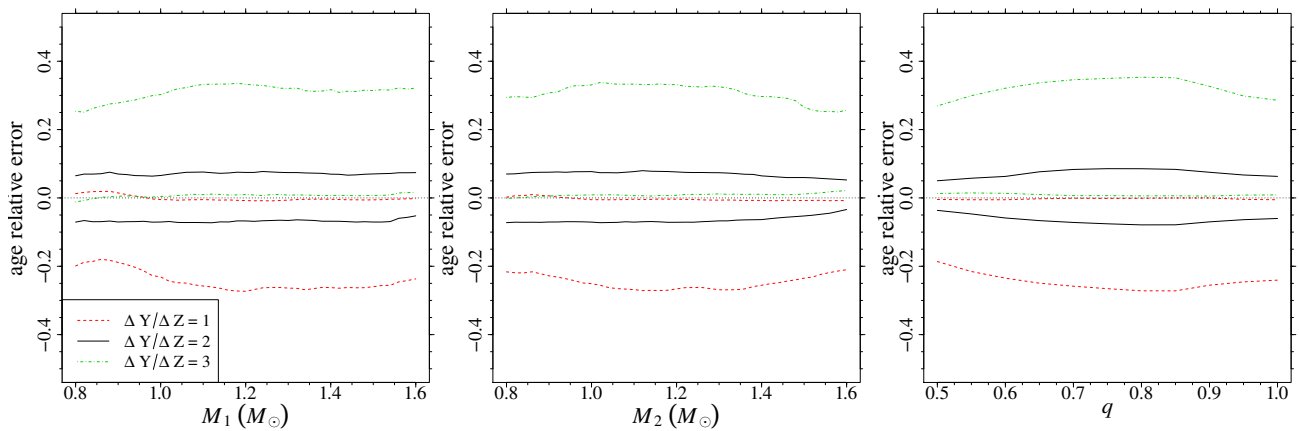}
\caption{{\it Left}: standard $1 \sigma$ envelope of the binary system age
  relative error in 
  dependence on the mass of the primary star (solid black), compared with the
  same quantity obtained by sampling from the grid with a different initial helium
  abundance computed assuming $\Delta Y/\Delta Z = 1$ (dashed red) and $\Delta
  Y/\Delta Z = 3$ (dot
  dashed green).
{\it Middle}: same as the {\it left panel}, but in dependence on the
mass of the secondary star. {\it Right}: same as the {\it left panel}, but in
 dependence on the system  mass ratio $q$.}
\label{fig:error-He}
\end{figure*}

\begin{figure*}
\centering
\includegraphics[height=8.5cm,angle=-90]{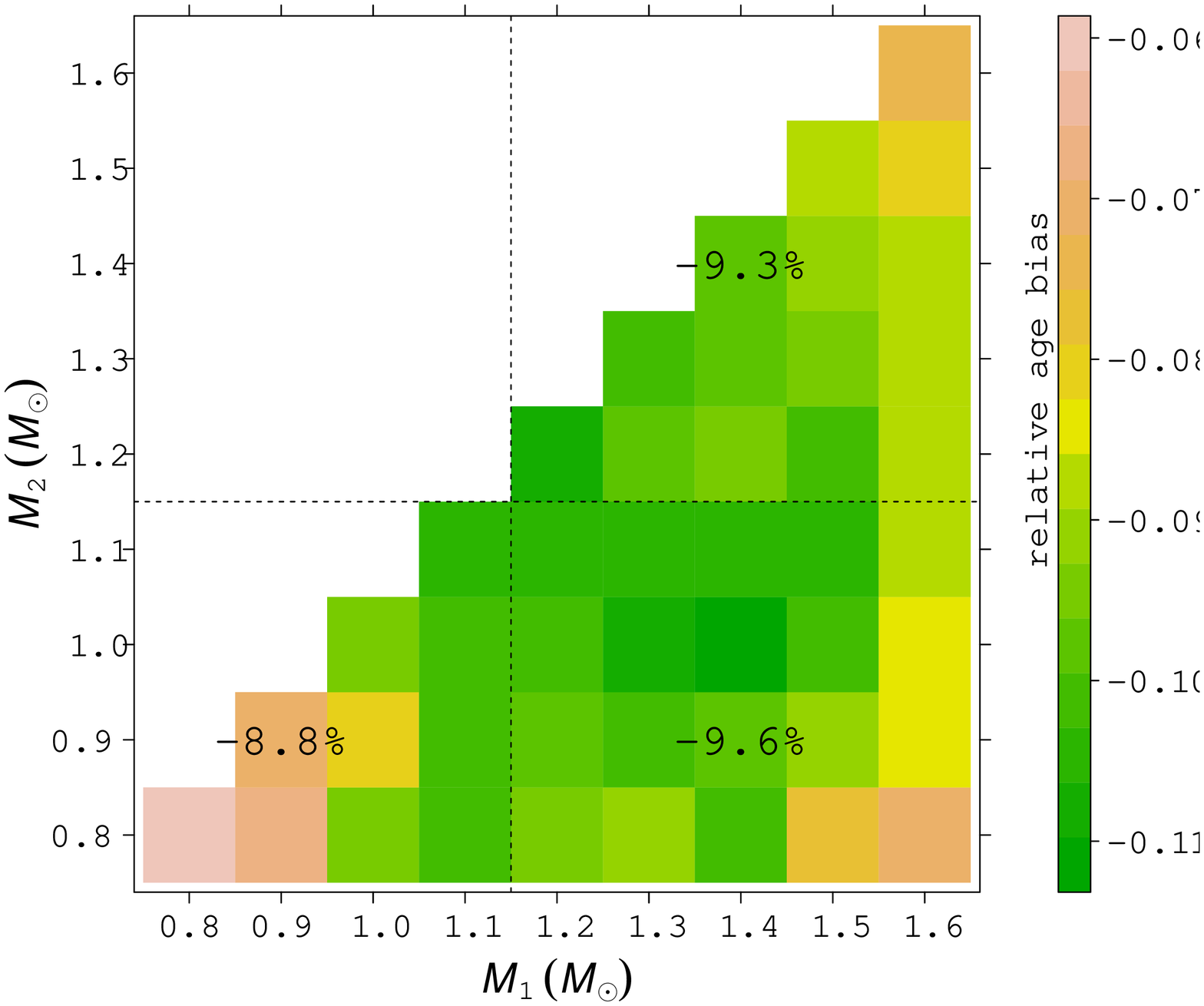}
\includegraphics[height=8.5cm,angle=-90]{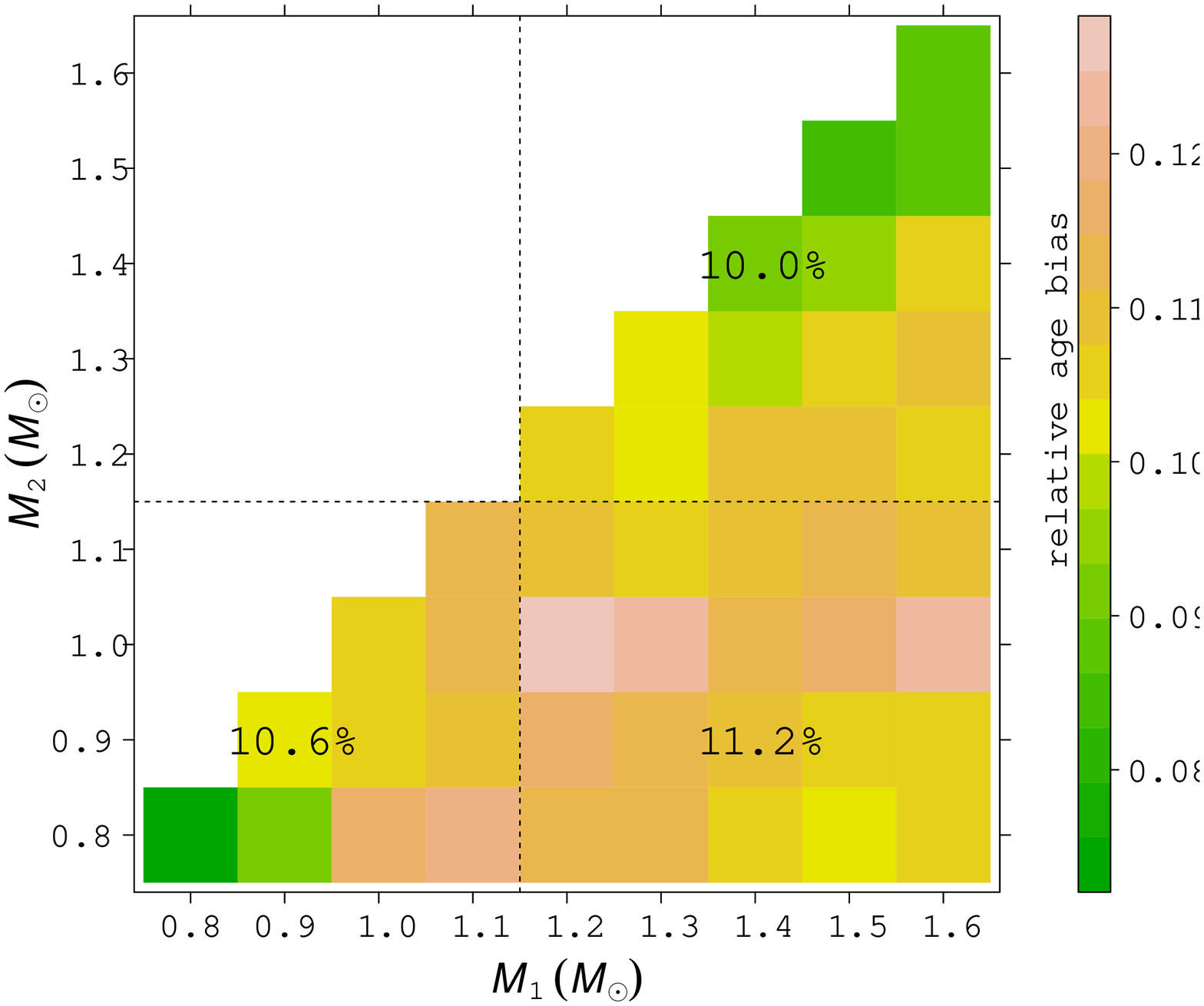}
\caption{{\it Left}: relative-age bias due to adopting a low initial helium abundance
        as a function of the masses of the binary system. Artificial stars      
        are sampled from the grid with  $\Delta Y/\Delta Z = 1,$ and their age  
        is estimated on the standard grid ($\Delta Y/\Delta Z = 2$). {\it Right}: same as the
        {\it left panel}, but sampling from a grid
        with  $\Delta Y/\Delta Z = 3$.}
\label{fig:envelop-M1_M2-diff-He}
\end{figure*}

The results are presented in Tables~\ref{tab:biasM} and \ref{tab:biasq} and in
Figs.~\ref{fig:error-He} and 
\ref{fig:envelop-M1_M2-diff-He}.
The mean impact of the initial helium abundance uncertainty is higher than
for the convective core overshooting.
The median bias for $\Delta Y/\Delta Z = 1$ -- computed as a function of the
mass ratio $q$ (Table ~\ref{tab:biasq}) -- is about the $-140\%$ 
of the half-width of the standard $1 \sigma$ 
envelope due to observational uncertainties reported in Table~\ref{tab:quantiles}. The corresponding bias for
$\Delta Y/\Delta Z = 3$ is about 160\% of the standard envelope
half width in Table~\ref{tab:quantiles}. 
Figure~\ref{fig:envelop-M1_M2-diff-He} shows that
both the changes in the bias due to the primary and secondary stars are modest, 
so, in the examined mass range, the initial helium induced bias can be considered almost constant at about $\pm 10\%$ for a change in $\Delta Y/\Delta Z$ by $\pm 1$. 

The results presented in this section on the initial helium uncertainty impact are very different from those reported in V15, where this uncertainty source have been found to be nearly negligible. This striking difference results 
from the observational constraints adopted in the estimation. The analysis in V15 in fact showed that the little influence of the helium content on age estimates came from two opposite effects that nearly cancel each other. The first obvious effect was due to the impact of the helium content on the evolutionary time scale, so that helium-rich models evolve faster. A second effect was due to the placement on the standard recovery grid of the non-standard helium models; helium-rich models mimicked more massive standard  models, leading to a mass overestimation bias and therefore to reduced age estimates. 
The two effects compensate for each other. 

This balancing effect cannot occur here since the mass is adopted as an observational constraint, and this leads to the strong influence of the initial helium content on the final age estimates.         

\subsection{Element microscopic diffusion}\label{sec:elements-microscopic-diffusion}

We evaluated the impact on the age estimates of neglecting 
microscopic diffusion in the stellar models adopted in the recovery procedure following, as in V15, a 
different approach with respect to the previous cases. Here we preferred to build the synthetic data set by sampling $N = 50\,000$ artificial 
binary systems from stellar models that take diffusion into account and by adding Gaussian noise. We then reconstructed  
 their age by means of the SCEPtER pipeline but adopting the non-standard grid of models, which neglect diffusion.

\begin{figure*}
\centering
\includegraphics[height=17cm,angle=-90]{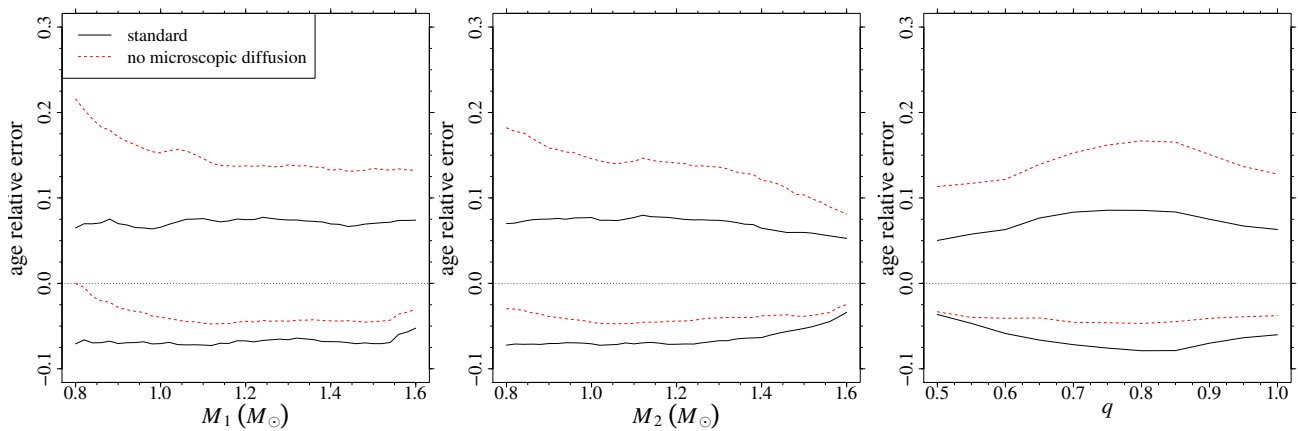}
\caption{{\it Left}: standard $1 \sigma$ envelope of the binary-system age
-relative error in 
  dependence on the mass of the primary star (solid black), compared with the
  same quantity obtained by sampling from the standard grid and reconstruction
  on a grid of stellar models computed without microscopic diffusion (dashed
  red). 
{\it Middle}: same as the {\it left panel}, but in dependence on the
mass of the secondary star. {\it Right}: same as the {\it left panel}, but in dependence on
  the system's 
  mass ratio $q$.}
\label{fig:error-noFeH}
\end{figure*}

\begin{figure}
\centering
\includegraphics[height=8.5cm,angle=-90]{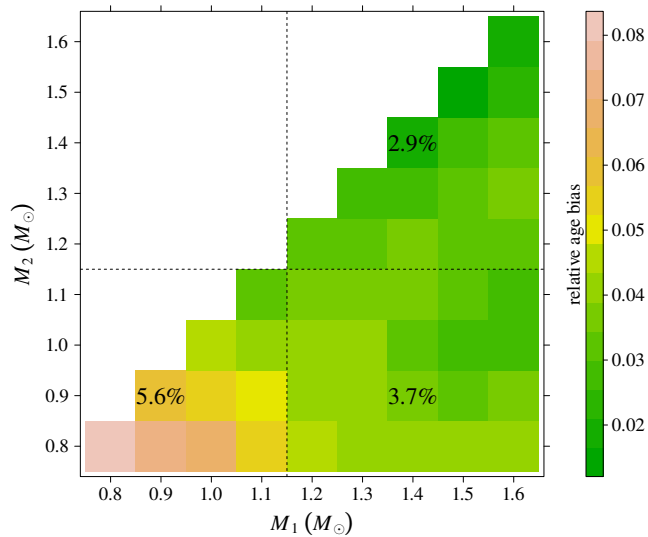}
\caption{Relative-age bias due to neglecting elements diffusion
        as a function of the masses of the binary system. Artificial stars      
        are sampled from the standard grid, and their ages are
estimated on a grid computed without microscopic diffusion.}
\label{fig:envelop-M1_M2-diff-noFeH}
\end{figure}

The results are presented in Tables~\ref{tab:biasM} and \ref{tab:biasq} and in
Figs.~\ref{fig:error-noFeH} and 
\ref{fig:envelop-M1_M2-diff-noFeH}. The bias due to the microscopic diffusion
 has a mean value of 3.6\% (see  Table~\ref{tab:biasq}), it is 
similar in absolute value to the one due to mild overshooting, and it ranges from about one-half to one-third of the bias due to initial helium change. 
Figure~\ref{fig:envelop-M1_M2-diff-noFeH} confirms an expected result that the diffusion is most important in systems composed of two low-mass stars. For primary and secondary masses lower than 1.1 $M_{\sun}$, the bias in age estimates is 5.6\%, while it drops to 3.7\% for more massive primary stars. 
The trend can be understood since the evolutionary time scale of these stars is faster than for the diffusion processes, which are therefore quite inefficient (see left panel in
Fig.~\ref{fig:error-noFeH}).

As already discussed in Sect.~\ref{sec:he}, this result confirms that the bias in age grid estimates, owing to the different  input in the evolutionary stellar codes, and strongly depends on the assumed observational constraints. In fact V15 has shown that -- if adopting asteroseismic constraints -- the microscopic diffusion-induced bias dominated the other examined biases, while it is not the case here.

\section{The coevality problem: a statistical approach}
\label{sec:expected-diff}

In the literature there are several cases of real eclipsing binary systems whose components 
are estimated to be non-coeval, because they cannot be fitted simultaneously by a single isochrone \citep[e.g.][]{Clausen2009,Vos2012,Sandquist2013}. Such an occurrence 
is often used to claim that the current generation of stellar models present some weaknesses and that additional 
physical processes should be included or re-calibrated. Given its considerable implications, it is worth discussing 
whether the estimates of non-coevality amongst the two components of real binary systems are statistically reliable or are simply 
a fluctuation. In this regard it is of particular interest to quantify the expected difference in the estimated ages of two genuine coeval stars
 caused simply by the observational uncertainties. To do that, we analysed our previously described synthetic 
 datasets of coeval binary stars.

As detailed in Sect.~\ref{sec:method}, the artificial binary systems sampled from the
grid of stellar models are subject to random perturbations to account for the
observational errors. Therefore the ages of the single stellar members, which are 
equal by construction in the sampling stage, are generally estimated as different. 
It is interesting to estimate how large the expected
differences are in age owing only to these random perturbations, since this effect
also interests real world observations. 

A detailed analysis of the problem has to deal with many technical aspects\footnote{Such
as the choice of the sampling scheme for the binary systems, the evaluation of the
precision and the accuracy on the estimated critical values, and of 
the sample size needed to attain a given level
of precision.} and it is  
therefore outside the aims of the present work. A devoted paper  
addresses the statistical aspects of this question presenting more detailed results (Valle et al. 2015, in prep.).    
Nevertheless, we present some basic findings here  that were obtained with the standard
sampling  described in Sect.~\ref{sec:method}. 

We define $A_1$ and $A_2$ as the estimated ages of the two members, with $A_1 > A_2$.
We focused our attention on the statistics $W = (A_1-A_2)/A_1$, which
then ranges from 0, when the stars are correctly estimated coeval,
and 1, when one star is estimated to be much younger than the other. 
To develop a statistical test based on $W$ we have to estimate how large $W$ 
can become only because of the random fluctuations in age estimates $A_1$ and $A_2$.
Once the 
distribution of $W$ -- by means of a Monte Carlo simulation has been empirically estimated, it is possible to choose a critical value by 
identifying the range of values of $W$ that are too extreme to be compatible with the coevality hypothesis.
Usually the critical value is chosen as the 95th or the 99th quantile of the 
distribution of the statistics under consideration. The choice of the 
critical value to adopt (in the following, we refer to this as  $W_{1-\alpha}$)  defines 
the "level" $\alpha$ of the test ($\alpha$ thus being the probability that an observed value of 
$W$ is larger than $W_{1-\alpha}$). The set of values 
higher than the critical values are said to lie in the "rejection region".

We present the rejection regions at level $\alpha$ = 0.05, hence the critical values $W_{0.95}$ identifying them, of the
null hypothesis that 
the stars are coeval. Reconstructed values of $W$ higher than the critical
value $W_{0.95}$ lead to the rejection of the null hypothesis, implying that standard stellar models
 cannot account for the coevality of the stars with the assumed
input or parameters.

A first obvious complication comes from the value of $W_{0.95}$
depending on the mass of the two stars, on their metallicity, and on the
(observationally unknown) relative age. Here we focus on the impact of the binary star masses. 

To perform the critical-value estimation, we generated a
sample of $N = 50\,000$ artificial binary systems, perturbed their observables as explained in Sect.~\ref{sec:method}, and then independently estimated the ages of the two
stars. From these estimates, we computed the statistics $W$. Then we binned the
$W$ values according to the primary and secondary stellar masses, as for constructing the 2D envelope in Sect.~\ref{sec:perturbation}.
In each of these bins, we computed the empirical $W$ 95th quantile, which
approximates the required critical value. A detailed discussion of the dependence of 
such a critical value on the actual uncertainties affecting the observational constraints and on the 
evolutionary phase of the two binary members will be presented in a forthcoming paper (Valle et al. 2015, in prep.).

\begin{figure}
\centering
\includegraphics[height=8.5cm,angle=-90]{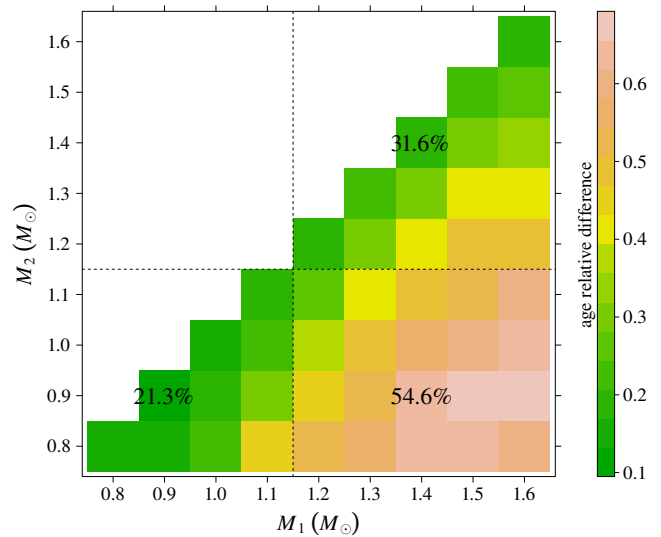}
\caption{Critical values $W_{0.95}$ of the statistic $W$ for the
  expected differences in the ages of the two stars, only due to the
  observational uncertainties (see text).}
\label{fig:envelop-M1_M2-diffage}
\end{figure}

The results of the analysis are displayed in
 Fig.~\ref{fig:envelop-M1_M2-diffage}. The figure shows
 that genuine coeval stars can be reconstructed as non-coeval with a sizeable 
 relative age difference only because of the current 
 uncertainty in the observational constraints. 
While the critical value $W_{0.95}$ is about
0.25 for systems of nearly equal masses, it can be greater than 0.6 for unbalanced binary masses. 

This result should be carefully considered  when determining the
stellar ages of a binary system. 
Thus, a determination of two discrepant stellar ages, which leads to a value of $W < W_{0.95}$, does not allow a statistically grounded rejection of the 
hypothesis that the stars are coeval since the difference might only be due to a random
fluctuation in the observational constraints.  

As discussed in Sect.~\ref{sec:ov}, one should be aware of this effect whenever the extent of convective core overshooting is 
calibrated on eclipsing binary by requiring that both stellar components are simultaneously fitted by a single isochrone, i.e. by imposing their coevality. If the ages estimated 
by means of stellar models without core overshooting give a value of  $W < W_{0.95}$, one cannot conclude on statistical grounds that these models are not able to simultaneously 
fit both stars and that a convective core overshooting must be taken into account. Therefore,
the reliability of the calibration method of the convective core overshooting consisting in the isochrones fine tuning to the observation has to be cautiously evaluated. 

In contrast, when standard stellar models provide ages of the binary components such that $W > W_{0.95}$, the possibility that non-coevality is the result of observational 
uncertainties should be considered as negligible. As an example, binary systems in which at least one star is suspected of supporting a strong surface activity, generally 
show a large discrepancy among the ages of the two components, such as V636 Cent and EF Aqr \citep{Clausen2009, Vos2012}. In this system, the active secondary stars are estimated to be much older than the primary, far above the limit due to random fluctuations. 
The coevality is recovered adopting a significantly lower mixing-length value for the secondary stars. This is an expected behaviour since it is recognised that in active stars the dynamo-induced magnetic fields can suppress
convection and produce starspots, causing differences in temperature and radii with
respect to standard stellar models, which can be mimicked by a lower mixing-length value \citep{Gabriel1969,Cox1981,Clausen1999}.  
It is, however, clear that further research is mandatory to determine the expected errors in the calibrated mixing length in
a sound statistical way.

\section{Conclusions}\label{sec:conclusions}

We performed a detailed theoretical investigation of the age estimation 
of detached double-lined eclipsing binaries by means of a maximum likelihood
grid-based technique. 
We analysed the impact of the current uncertainty on several input of stellar
model computations, such as the chemical composition, the efficiency of
convective core overshooting, and the efficiency of microscopic processes.

We adopted the grid-based pipeline SCEPtER, which is extensively described in
\citet{scepter1} and \citet{eta}. As
observational constraints we assumed the stellar effective temperature, the
metallicity 
[Fe/H], the mass, and the radius of the stars. The grid of stellar models,
computed for the evolutionary phases from ZAMS to the central hydrogen
depletion  covers 
the mass range [0.8; 1.6] $M_{\sun}$.

We computed the statistical errors arising
from the uncertainties in observational quantities. We found an overall
relative uncertainty of about $\pm 7\%$ in  
age estimates of the system, which is nearly independent of the masses of the stars and
of the mass ratio of the binary systems. However, as described in \citet{eta}, the relative error in the age estimates 
sensitively depends on the evolutionary phase and becomes larger for models near those ZAMS
 for which it can reach values of about 90\% for the upper boundary, with an envelope half width of about 50\%. 
The large uncertainty and the bias toward higher ages found near the ZAMS are mainly an edge effect distortion, since the estimated age cannot be negative.
 The 
relative error envelope quickly shrinks as the stars evolve from the ZAMS, 
and for primary stars 
of relative age 0.3, the half width of the error envelope is about 15\%. 

We studied the impact on the age estimates of different choices of
the sampling algorithms, of the correlation between 
the effective temperatures, the masses, the radii, and [Fe/H] of the two binary components, and of the method of estimating the
joint age of the stars.
It resulted that the first source of uncertainty
plays a minor role since it accounts for a variation in the relative error
half width of about 1\%.  The neglect of correlations between the observables of the two stars was shown to reduce the age error envelope width by about one-fifth.
The age estimates of the binary
system obtained by averaging the ages of the two stars computed independently
-- without explicitly assuming coevality --  lead to negligible differences only for stars
of nearly equal masses. Whenever an unbalanced system is considered ($M_1 > 1.1$ $M_{\sun}$ and $M_2 \leq 1.1$ $M_{\sun}$), the
envelope width of the age relative errors for the average of independent age estimates is larger by a factor of two, but it
can be as high as a factor of five  for systems with $q \approx 0.5$.

We evaluated the
systematic biases on age estimates owing to the uncertainties in initial helium
content,  in the 
convective core overshooting, and in the 
microscopic diffusion efficiency. 
We found that the bias due to a mild ($\beta$ = 0.2) and strong ($\beta$ = 0.4)  convective core overshooting
scenarios are about 50\% to 120\% of the half width of the standard
age relative error envelope due only to the observational uncertainties.
The bias due to the initial helium content was explored assuming an
uncertainty of $\pm 1$ on the helium-to-metal enrichment ratio $\Delta
Y/\Delta Z$. The effect is about $-140\%$
and 160\% of the $1 \sigma$ random error, so slightly more than that of the strong overshooting scenario.
Finally, the bias due to the neglect of the element microscopic
diffusion was about 60\% of the half width of the standard
age-relative error envelope. 

The comparison of these results to those reported in \citet{eta}, which were computed by adopting asteroseismic constraints and without the knowledge of stellar mass and radius, showed relevant differences. In fact, \citet{eta} showed that the helium-induced bias on age estimates was negligible, while the bias due to the microscopic diffusion was dominant. These differences arise from the corresponding bias on mass estimates that could occur in \citet{eta}, which led to some compensations for the age estimates. These mechanisms cannot occur in the present study, since the stellar mass is assumed as an observational constraint. An important point is therefore that the bias in the grid-based age estimates -- due to adopting different input in a stellar evolutionary code -- strongly depends on the assumed observational constraints and should be evaluated case-by-case whenever a new grid is adopted for estimates. 

We also introduced a statistical test on the expected differences in the
estimated ages of two genuine coeval binary components due simply to the 
observational uncertainties affecting the observables used in the recovery procedure.
We found that the relative difference between the reconstructed ages of two coeval binary members depends on the 
mass ratio $q$. For systems with $q \approx 1$, the relative difference in the 
inferred ages can be about 20\%, while it is 
higher than 60\% for system with $q \lesssim 0.8$.
This result should be carefully taken into account before 
using the apparent non-coevality of stars in binary systems -- i.e. the inability to find a single isochrone that fits both components -- 
to claim the weakness of current generation of stellar models and to 
calibrate the efficiency of some poorly known physical process, such as convective core overshooting, by isochrone fine tuning with the observations.

\begin{acknowledgements}
We thank our anonymous referee for the stimulating comments that helped us to clarify and improve the paper. 
This work has been supported by PRIN-MIUR 2010-2011 ({\em Chemical and dynamical evolution 
        of the Milky Way and Local Group galaxies}, PI F. Matteucci), and PRIN-INAF 2012 
({\em The M4 Core Project with Hubble Space Telescope}, PI
L. Bedin).      
\end{acknowledgements}

\appendix

\section{Synthetic dataset: sampling strategy}\label{app:sampling} 

In constructing the sample adopted in the error estimation, we
did not try to match any observed distribution of the mass of the
  binary stars or of the mass ratio $q$. 
  Since we primarily aimed to investigate the performances 
  of grid-based age estimates of binary stars, we paid more
attention when sampling the whole range in mass, metallicity, and  evolutionary phases covered 
by the available grid of models. 

In detail, the datasets of artificial stars are sampled by adopting the
following scheme. First, 
a star is randomly sampled from the grid. No priors on mass, age, and metallicity are assumed
in the sampling procedure. Then, a second star is coupled to the first
one with 
two requirements: it must share with the first star both the same initial
 [Fe/H] and the age, with a tolerance of 10 Myr. The couple of stars is re-ordered to have the most luminous star as a
first member. 

\begin{figure*}
\centering
\includegraphics[height=17cm,angle=-90]{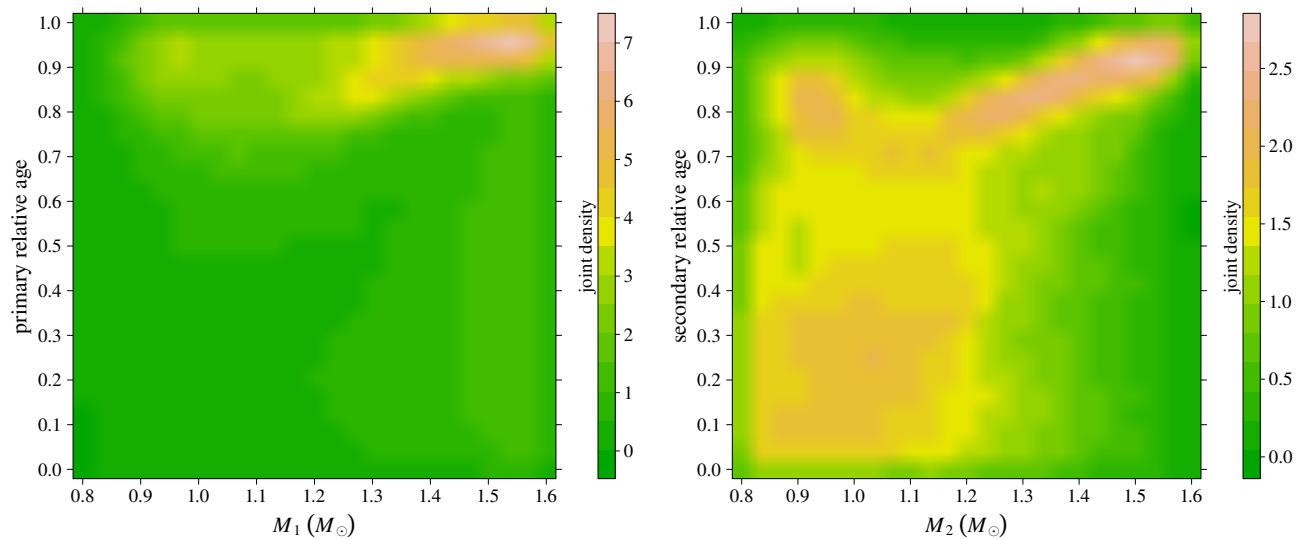}
\caption{{\it Left}: joint density of mass and relative ages in the sample of
  primary stars. The colours correspond to the different densities of
  probability. {\it Right}: same as the {\it left panel} for secondary 
  stars. The colour scales
of the two panels are different.}
\label{fig:density-M_pcage}
\end{figure*}

\begin{figure}
\centering
\includegraphics[height=8cm,angle=-90]{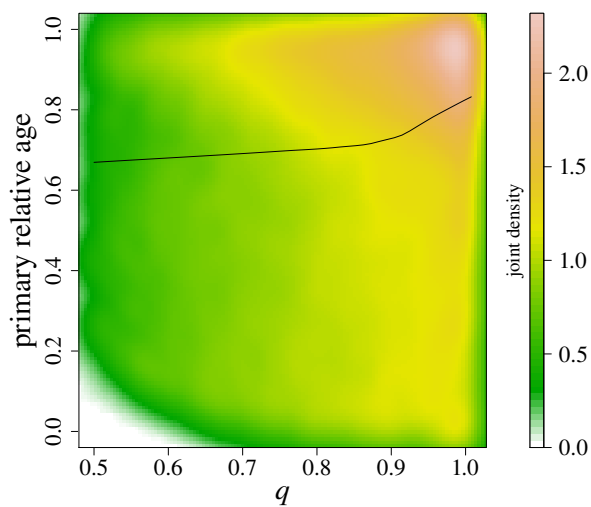}
\caption{Joint density of mass ratio $q$ and primary relative age in the sample of
  binary stars. The solid line displays the LOESS-smoothed trend of
  relative age 
  versus mass ratio $q$.}
\label{fig:density-q_pcage}
\end{figure}

To show which models are selected as primary and secondary stars, we estimated
the joint density of mass and relative age for primary and secondary stars\footnote{
The joint density was obtained by a bi-dimensional kernel density 
estimation using the function {\it kde2d} in the R library MASS. Details on the technique can be found 
in \citet{venables2002modern}.}.
From the joint density it is possible to verify if there are ranges of mass or
relative age that are preferably sampled.
The results are displayed in Figure~\ref{fig:density-M_pcage}. It is apparent
that the density functions for the primary and 
secondary stars are different and that they are both not uniform.
In particular, the sample of primary stars is biased towards high
masses and high 
relative ages since these models have higher intrinsic luminosity. The
distribution of secondary stars is more diffuse, since several young low-mass
models are present in the sample. 
Figure~\ref{fig:density-q_pcage} shows the joint density of the mass ratio $q$
and  the primary
relative age. The solid line in the figure is a {\it LOESS} smoother\footnote{A LOESS 
(LOcal regrESSion) smoother is 
a non-parametric,  locally weighted polynomial regression technique that is
often used to show the underlying trend of scattered data \citep[see
e.g.][]{Feigelson2012,venables2002modern}.}
of the relative age versus $q$. 
It is apparent that for binary systems with
near equal masses, the sampling returns more evolved stars. This bias comes from later evolutionary phases  needing more points to adequately follow the rapid evolution. The median time step in this grid region is about 13 Myr.
Once a model has been sampled in this grid region, several points in the same evolutionary track will pass the constraint on maximum age differences of 10 Myr from the first sampled model. 
Therefore there
is a high probability that a point in the same track is coupled to the first selected stars, leading to a mass ratio of 1.0. This phenomenon is not important in the reconstruction phase,
 since all these points have basically the same age, as discussed in V15. 

The effect of the sampling could be
seen in Table~\ref{tab:quantiles}, which shows a small
shrink in the age relative 
error envelope at $q$ greater than 0.9. This is a direct consequence of
the relative error envelope narrowing at high relative age. 
To show that this artificial trend could be removed with different sampling strategies,
in Table~\ref{tab:quantiles} we also present two additional series of results. 
In fact, we first explored the
impact of a different sampling. For this test, we sampled the set of
primary stars, then we coupled each of those objects to the secondary star by sampling
only from the models respecting the constraint in metallicity and age, which are
also less massive than the primary object. The results are in the rows
labelled ``mass ratio $q$ (alternative sampling)'' in
Table~\ref{tab:quantiles}. As a second test, to see the impact of the
ad-hoc removal of the trend of primary relative ages versus $q$ shown in
 Fig.~\ref{fig:density-q_pcage}, we checked an alternative approach 
that rejects some objects in the subset $q >
0.9$, based on the standard method. In fact, since this is the region that is more populated, it was possible to draw from it
a sample of size equal to that of the adjacent subset $0.8
< q \leq 0.9$, with the requirement that the new sample's primary relative age
matches the distribution of the subset $0.8
< q \leq 0.9$. The corresponding results are in the rows  "mass ratio $q$
(rejection step)" in
Table~\ref{tab:quantiles}. It appears that the shrinkage in the upper $q$ region
is reduced in both two cases.

In contrast, the shrinkage of the error envelope at low mass ratio $q$ (see Table~\ref{tab:quantiles}) is due to an edge effect since there are only models in 
this region with mass 1.6 $M_{\sun}$ coupled with 0.8
$M_{\sun}$ models. In fact the lower left-hand region of Fig.~\ref{fig:density-q_pcage} is not populated because  
to produce a coeval combination, the more
massive model cannot be too young. In fact, in this case the secondary is still in the pre-main sequence phase, which is not considered in the grid.   Figure~\ref{fig:density-q_pcage}
shows that the minimum relative age for these massive models is about
0.2. Therefore the most critical region for age estimation -- that of low
relative age -- is avoided, and the estimate relative error envelope is
narrower.

\bibliographystyle{aa}
\bibliography{biblio}

\begin{thebibliography}{60}
\expandafter\ifx\csname natexlab\endcsname\relax\def\natexlab#1{#1}\fi

\bibitem[{{Andersen}(1991)}]{Andersen1991}
{Andersen}, J. 1991, \aapr, 3, 91

\bibitem[{{Asplund} {et~al.}(2009){Asplund}, {Grevesse}, {Sauval}, \&
  {Scott}}]{AGSS09}
{Asplund}, M., {Grevesse}, N., {Sauval}, A.~J., \& {Scott}, P. 2009, \araa, 47,
  481

\bibitem[{{Basu} {et~al.}(2010){Basu}, {Chaplin}, \& {Elsworth}}]{Basu2010}
{Basu}, S., {Chaplin}, W.~J., \& {Elsworth}, Y. 2010, \apj, 710, 1596

\bibitem[{{Basu} {et~al.}(2012){Basu}, {Verner}, {Chaplin}, \&
  {Elsworth}}]{Basu2012}
{Basu}, S., {Verner}, G.~A., {Chaplin}, W.~J., \& {Elsworth}, Y. 2012, \apj,
  746, 76

\bibitem[{{Bonaca} {et~al.}(2012){Bonaca}, {Tanner}, {Basu}, {Chaplin},
  {Metcalfe}, {Monteiro}, {Ballot}, {Bedding}, {Bonanno}, {Broomhall},
  {Bruntt}, {Campante}, {Christensen-Dalsgaard}, {Corsaro}, {Elsworth},
  {Garc{\'{\i}}a}, {Hekker}, {Karoff}, {Kjeldsen}, {Mathur}, {R{\'e}gulo},
  {Roxburgh}, {Stello}, {Trampedach}, {Barclay}, {Burke}, \&
  {Caldwell}}]{Bonaca2012}
{Bonaca}, A., {Tanner}, J.~D., {Basu}, S., {et~al.} 2012, \apjl, 755, L12

\bibitem[{{Brogaard} {et~al.}(2011){Brogaard}, {Bruntt}, {Grundahl}, {Clausen},
  {Frandsen}, {Vandenberg}, \& {Bedin}}]{Brogaard2011}
{Brogaard}, K., {Bruntt}, H., {Grundahl}, F., {et~al.} 2011, \aap, 525, A2

\bibitem[{{Claret}(2003)}]{Claret2003}
{Claret}, A. 2003, \aap, 399, 1115

\bibitem[{{Claret}(2007)}]{Claret2007}
{Claret}, A. 2007, \aap, 475, 1019

\bibitem[{{Clausen} {et~al.}(2009){Clausen}, {Bruntt}, {Claret}, {Larsen},
  {Andersen}, {Nordstr{\"o}m}, \& {Gim{\'e}nez}}]{Clausen2009}
{Clausen}, J.~V., {Bruntt}, H., {Claret}, A., {et~al.} 2009, \aap, 502, 253

\bibitem[{{Clausen} {et~al.}(2010){Clausen}, {Frandsen}, {Bruntt}, {Olsen},
  {Helt}, {Gregersen}, {Juncher}, \& {Krogstrup}}]{Clausen2010}
{Clausen}, J.~V., {Frandsen}, S., {Bruntt}, H., {et~al.} 2010, \aap, 516, A42

\bibitem[{{Clausen} {et~al.}(1999){Clausen}, {Helt}, \& {Olsen}}]{Clausen1999}
{Clausen}, J.~V., {Helt}, B.~E., \& {Olsen}, E.~H. 1999, in Astronomical
  Society of the Pacific Conference Series, Vol. 173, Stellar Structure: Theory
  and Test of Connective Energy Transport, ed. A.~{Gimenez}, E.~F. {Guinan}, \&
  B.~{Montesinos}, 321

\bibitem[{{Clausen} {et~al.}(2008){Clausen}, {Torres}, {Bruntt}, {Andersen},
  {Nordstr{\"o}m}, {Stefanik}, {Latham}, \& {Southworth}}]{Clausen2008}
{Clausen}, J.~V., {Torres}, G., {Bruntt}, H., {et~al.} 2008, \aap, 487, 1095

\bibitem[{{Cox} {et~al.}(1981){Cox}, {Hodson}, \& {Shaviv}}]{Cox1981}
{Cox}, A.~N., {Hodson}, S.~W., \& {Shaviv}, G. 1981, \apjl, 245, L37

\bibitem[{{Cyburt} {et~al.}(2004){Cyburt}, {Fields}, \& {Olive}}]{cyburt04}
{Cyburt}, R.~H., {Fields}, B.~D., \& {Olive}, K.~A. 2004, \prd, 69, 123519

\bibitem[{{Degl'Innocenti} {et~al.}(2008){Degl'Innocenti}, {Prada Moroni},
  {Marconi}, \& {Ruoppo}}]{scilla2008}
{Degl'Innocenti}, S., {Prada Moroni}, P.~G., {Marconi}, M., \& {Ruoppo}, A.
  2008, \apss, 316, 25

\bibitem[{{Deheuvels} \& {Michel}(2011)}]{Deheuvels2011}
{Deheuvels}, S. \& {Michel}, E. 2011, \aap, 535, A91

\bibitem[{Dell'Omodarme \& Valle(2013)}]{stellar}
Dell'Omodarme, M. \& Valle, G. 2013, The R Journal, 5, 108

\bibitem[{{Dell'Omodarme} {et~al.}(2012){Dell'Omodarme}, {Valle},
  {Degl'Innocenti}, \& {Prada Moroni}}]{database2012}
{Dell'Omodarme}, M., {Valle}, G., {Degl'Innocenti}, S., \& {Prada Moroni},
  P.~G. 2012, A\&A, 540, A26

\bibitem[{Feigelson \& Babu(2012)}]{Feigelson2012}
Feigelson, E.~D. \& Babu, G.~J. 2012, Modern Statistical Methods for Astronomy
  with R applications (Cambridge University Press)

\bibitem[{{Gabriel}(1969)}]{Gabriel1969}
{Gabriel}, M. 1969, in Low-Luminosity Stars, ed. S.~S. {Kumar}, 267

\bibitem[{{Gai} {et~al.}(2011){Gai}, {Basu}, {Chaplin}, \&
  {Elsworth}}]{Gai2011}
{Gai}, N., {Basu}, S., {Chaplin}, W.~J., \& {Elsworth}, Y. 2011, \apj, 730, 63

\bibitem[{{Gennaro} {et~al.}(2010){Gennaro}, {Prada Moroni}, \&
  {Degl'Innocenti}}]{gennaro10}
{Gennaro}, M., {Prada Moroni}, P.~G., \& {Degl'Innocenti}, S. 2010, \aap, 518,
  A13+

\bibitem[{{Gennaro} {et~al.}(2012){Gennaro}, {Prada Moroni}, \&
  {Tognelli}}]{Gennaro2012}
{Gennaro}, M., {Prada Moroni}, P.~G., \& {Tognelli}, E. 2012, \mnras, 420, 986

\bibitem[{{Grundahl} {et~al.}(2008){Grundahl}, {Clausen}, {Hardis}, \&
  {Frandsen}}]{Grundahl2008}
{Grundahl}, F., {Clausen}, J.~V., {Hardis}, S., \& {Frandsen}, S. 2008, \aap,
  492, 171

\bibitem[{{He{\l}miniak} {et~al.}(2009){He{\l}miniak}, {Konacki}, {Ratajczak},
  \& {Muterspaugh}}]{Helminiak2009}
{He{\l}miniak}, K.~G., {Konacki}, M., {Ratajczak}, M., \& {Muterspaugh}, M.~W.
  2009, \mnras, 400, 969

\bibitem[{{Jimenez} {et~al.}(2003){Jimenez}, {Flynn}, {MacDonald}, \&
  {Gibson}}]{jimenez03}
{Jimenez}, R., {Flynn}, C., {MacDonald}, J., \& {Gibson}, B.~K. 2003, Science,
  299, 1552

\bibitem[{{Lacy} {et~al.}(2008){Lacy}, {Torres}, \& {Claret}}]{Lacy2008}
{Lacy}, C.~H.~S., {Torres}, G., \& {Claret}, A. 2008, \aj, 135, 1757

\bibitem[{{Lastennet} \& {Valls-Gabaud}(2002)}]{Lastennet2002}
{Lastennet}, E. \& {Valls-Gabaud}, D. 2002, \aap, 396, 551

\bibitem[{{Mathur} {et~al.}(2012){Mathur}, {Metcalfe}, {Woitaszek}, {Bruntt},
  {Verner}, {Christensen-Dalsgaard}, {Creevey}, {Do{\v g}an}, {Basu}, {Karoff},
  {Stello}, {Appourchaux}, {Campante}, {Chaplin}, {Garc{\'{\i}}a}, {Bedding},
  {Benomar}, {Bonanno}, {Deheuvels}, {Elsworth}, {Gaulme}, {Guzik}, {Handberg},
  {Hekker}, {Herzberg}, {Monteiro}, {Piau}, {Quirion}, {R{\'e}gulo}, {Roth},
  {Salabert}, {Serenelli}, {Thompson}, {Trampedach}, {White}, {Ballot},
  {Brand{\~a}o}, {Molenda-{\.Z}akowicz}, {Kjeldsen}, {Twicken}, {Uddin}, \&
  {Wohler}}]{Mathur2012}
{Mathur}, S., {Metcalfe}, T.~S., {Woitaszek}, M., {et~al.} 2012, \apj, 749, 152

\bibitem[{{Meibom} {et~al.}(2009){Meibom}, {Grundahl}, {Clausen}, {Mathieu},
  {Frandsen}, {Pigulski}, {Narwid}, {Steslicki}, \& {Lefever}}]{Meibom2009}
{Meibom}, S., {Grundahl}, F., {Clausen}, J.~V., {et~al.} 2009, \aj, 137, 5086

\bibitem[{{Morales} {et~al.}(2009){Morales}, {Torres}, {Marschall}, \&
  {Brehm}}]{Morales2009}
{Morales}, J.~C., {Torres}, G., {Marschall}, L.~A., \& {Brehm}, W. 2009, \apj,
  707, 671

\bibitem[{{Pagel} \& {Portinari}(1998)}]{pagel98}
{Pagel}, B.~E.~J. \& {Portinari}, L. 1998, \mnras, 298, 747

\bibitem[{{Pavlovski} {et~al.}(2014){Pavlovski}, {Southworth}, {Kolbas}, \&
  {Smalley}}]{Pavlovski2014}
{Pavlovski}, K., {Southworth}, J., {Kolbas}, V., \& {Smalley}, B. 2014, \mnras,
  438, 590

\bibitem[{{Peimbert} {et~al.}(2007{\natexlab{a}}){Peimbert}, {Luridiana}, \&
  {Peimbert}}]{peimbert07a}
{Peimbert}, M., {Luridiana}, V., \& {Peimbert}, A. 2007{\natexlab{a}}, \apj,
  666, 636

\bibitem[{{Peimbert} {et~al.}(2007{\natexlab{b}}){Peimbert}, {Luridiana},
  {Peimbert}, \& {Carigi}}]{peimbert07b}
{Peimbert}, M., {Luridiana}, V., {Peimbert}, A., \& {Carigi}, L.
  2007{\natexlab{b}}, in Astronomical Society of the Pacific Conference Series,
  Vol. 374, From Stars to Galaxies: Building the Pieces to Build Up the
  Universe, ed. {A.~Vallenari, R.~Tantalo, L.~Portinari, \& A.~Moretti}, 81--+

\bibitem[{{Pols} {et~al.}(1997){Pols}, {Tout}, {Schroder}, {Eggleton}, \&
  {Manners}}]{Pols1997}
{Pols}, O.~R., {Tout}, C.~A., {Schroder}, K.-P., {Eggleton}, P.~P., \&
  {Manners}, J. 1997, \mnras, 289, 869

\bibitem[{{Popper} {et~al.}(1986){Popper}, {Lacy}, {Frueh}, \&
  {Turner}}]{Popper1986}
{Popper}, D.~M., {Lacy}, C.~H., {Frueh}, M.~L., \& {Turner}, A.~E. 1986, \aj,
  91, 383

\bibitem[{{Prada Moroni} {et~al.}(2012){Prada Moroni}, {Gennaro}, {Bono},
  {Pietrzy{\'n}ski}, {Gieren}, {Pilecki}, {Graczyk}, \&
  {Thompson}}]{PradaMoroni2012}
{Prada Moroni}, P.~G., {Gennaro}, M., {Bono}, G., {et~al.} 2012, \apj, 749, 108

\bibitem[{{Quirion} {et~al.}(2010){Quirion}, {Christensen-Dalsgaard}, \&
  {Arentoft}}]{Quirion2010}
{Quirion}, P.-O., {Christensen-Dalsgaard}, J., \& {Arentoft}, T. 2010, \apj,
  725, 2176

\bibitem[{{Ribas} {et~al.}(2000){Ribas}, {Jordi}, \&
  {Gim{\'e}nez}}]{Ribas2000a}
{Ribas}, I., {Jordi}, C., \& {Gim{\'e}nez}, {\'A}. 2000, \mnras, 318, L55

\bibitem[{{Sandquist} {et~al.}(2013){Sandquist}, {Shetrone}, {Serio}, \&
  {Orosz}}]{Sandquist2013}
{Sandquist}, E.~L., {Shetrone}, M., {Serio}, A.~W., \& {Orosz}, J. 2013, \aj,
  146, 40

\bibitem[{{Schneider} {et~al.}(2014){Schneider}, {Langer}, {de Koter}, {Brott},
  {Izzard}, \& {Lau}}]{Schneider2014}
{Schneider}, F.~R.~N., {Langer}, N., {de Koter}, A., {et~al.} 2014, \aap, 570,
  A66

\bibitem[{{Southworth}(2013)}]{Southworth2013}
{Southworth}, J. 2013, \aap, 557, A119

\bibitem[{{Southworth} \& {Clausen}(2007)}]{Southworth2007}
{Southworth}, J. \& {Clausen}, J.~V. 2007, \aap, 461, 1077

\bibitem[{{Southworth} {et~al.}(2011){Southworth}, {Pavlovski}, {Tamajo},
  {Smalley}, {West}, \& {Anderson}}]{Southworth2011}
{Southworth}, J., {Pavlovski}, K., {Tamajo}, E., {et~al.} 2011, \mnras, 414,
  3740

\bibitem[{{Steigman}(2006)}]{steigman06}
{Steigman}, G. 2006, International Journal of Modern Physics E, 15, 1

\bibitem[{{Tanner} {et~al.}(2014){Tanner}, {Basu}, \& {Demarque}}]{Tanner2014}
{Tanner}, J.~D., {Basu}, S., \& {Demarque}, P. 2014, \apjl, 785, L13

\bibitem[{{Tognelli} {et~al.}(2011){Tognelli}, {Prada Moroni}, \&
  {Degl'Innocenti}}]{tognelli2011}
{Tognelli}, E., {Prada Moroni}, P.~G., \& {Degl'Innocenti}, S. 2011, \aap, 533,
  A109+

\bibitem[{{Torres} {et~al.}(2010){Torres}, {Andersen}, \&
  {Gim{\'e}nez}}]{Torres2010}
{Torres}, G., {Andersen}, J., \& {Gim{\'e}nez}, A. 2010, \aapr, 18, 67

\bibitem[{{Torres} {et~al.}(2012){Torres}, {Fischer}, {Sozzetti}, {Buchhave},
  {Winn}, {Holman}, \& {Carter}}]{Torres2012}
{Torres}, G., {Fischer}, D.~A., {Sozzetti}, A., {et~al.} 2012, \apj, 757, 161

\bibitem[{{Torres} {et~al.}(2006){Torres}, {Lacy}, {Marschall}, {Sheets}, \&
  {Mader}}]{Torres2006}
{Torres}, G., {Lacy}, C.~H., {Marschall}, L.~A., {Sheets}, H.~A., \& {Mader},
  J.~A. 2006, \apj, 640, 1018

\bibitem[{{Torres} {et~al.}(2009){Torres}, {Sandberg Lacy}, \&
  {Claret}}]{Torres2009}
{Torres}, G., {Sandberg Lacy}, C.~H., \& {Claret}, A. 2009, \aj, 138, 1622

\bibitem[{{Torres} {et~al.}(2014){Torres}, {Vaz}, {Sandberg Lacy}, \&
  {Claret}}]{Torres2014}
{Torres}, G., {Vaz}, L.~P.~R., {Sandberg Lacy}, C.~H., \& {Claret}, A. 2014,
  \aj, 147, 36

\bibitem[{{Valle} {et~al.}(2013{\natexlab{a}}){Valle}, {Dell'Omodarme}, {Prada
  Moroni}, \& {Degl'Innocenti}}]{incertezze1}
{Valle}, G., {Dell'Omodarme}, M., {Prada Moroni}, P.~G., \& {Degl'Innocenti},
  S. 2013{\natexlab{a}}, \aap, 549, A50

\bibitem[{{Valle} {et~al.}(2013{\natexlab{b}}){Valle}, {Dell'Omodarme}, {Prada
  Moroni}, \& {Degl'Innocenti}}]{incertezze2}
{Valle}, G., {Dell'Omodarme}, M., {Prada Moroni}, P.~G., \& {Degl'Innocenti},
  S. 2013{\natexlab{b}}, \aap, 554, A68

\bibitem[{{Valle} {et~al.}(2014){Valle}, {Dell'Omodarme}, {Prada Moroni}, \&
  {Degl'Innocenti}}]{scepter1}
{Valle}, G., {Dell'Omodarme}, M., {Prada Moroni}, P.~G., \& {Degl'Innocenti},
  S. 2014, \aap, 561, A125 (V14)

\bibitem[{{Valle} {et~al.}(2015){Valle}, {Dell'Omodarme}, {Prada Moroni}, \&
  {Degl'Innocenti}}]{eta}
{Valle}, G., {Dell'Omodarme}, M., {Prada Moroni}, P.~G., \& {Degl'Innocenti},
  S. 2015, \aap, 575, A12 (V15)

\bibitem[{{Valle} {et~al.}(2009){Valle}, {Marconi}, {Degl'Innocenti}, \& {Prada
  Moroni}}]{cefeidi}
{Valle}, G., {Marconi}, M., {Degl'Innocenti}, S., \& {Prada Moroni}, P.~G.
  2009, \aap, 507, 1541

\bibitem[{Venables \& Ripley(2002)}]{venables2002modern}
Venables, W. \& Ripley, B. 2002, Modern applied statistics with S, Statistics
  and computing (Springer)

\bibitem[{{Vos} {et~al.}(2012){Vos}, {Clausen}, {J{\o}rgensen}, {{\O}stensen},
  {Claret}, {Hillen}, \& {Exter}}]{Vos2012}
{Vos}, J., {Clausen}, J.~V., {J{\o}rgensen}, U.~G., {et~al.} 2012, \aap, 540,
  A64

\end{thebibliography}

\end{document}